\documentclass[preprint]{elsarticle}

\usepackage{cmap}
\usepackage{booktabs} % For formal tables

\usepackage{psfrag}
\usepackage{subfig}
\usepackage{graphicx}
\graphicspath{{./img/}}
\usepackage{epsfig}
\usepackage{enumitem}
\usepackage{tikz}
\usepackage{forloop}
\usepackage{mathtools}
\usepackage{amsmath}
\usepackage{amssymb}

\usepackage{lipsum}
\usepackage{amsfonts}
\usepackage{graphicx}
\usepackage{epstopdf}
\usepackage{algorithmic}

\journal{Communications in Nonlinear Science and Numerical Simulation}

\begin{document}

\begin{frontmatter}

\title{Effect of two forms of feedback on the performance of the Rate Control Protocol (RCP)}
%\title{Local Hopf Bifurcation Analysis of a Rate Control Protocol\,(RCP)}
%\title{Impact of queue size feedback on the performance of a Rate Control Protocol\,(RCP)}

\author[iitm]{Abuthahir\corref{mycorrespondingauthor}}
\ead{ee12d207@ee.iitm.ac.in}

\author[iitm]{Nizar Malangadan}
\ead{ee11s040@ee.iitm.ac.in}

\author[iitm]{Gaurav Raina}
\ead{gaurav@ee.iitm.ac.in}

\address[iitm]{Department of Electrical Engineering, Indian Institute of Technology Madras, Chennai-600036, India}
%\address[apple]{Independent Researcher}
%\address[apple]{Statistical Laboratory, University of Cambridge, Cambridge, UK}
\cortext[mycorrespondingauthor]{Corresponding author at: Department of Electrical Engineering, Indian Institute of Technology Madras, Chennai-600036, India. }

\begin{abstract}
The Rate Control Protocol (RCP) uses explicit feedback from routers to control network congestion. RCP estimates it's fair rate from two forms of feedback: rate mismatch and queue size. An important design question that remains open in RCP is whether the presence of queue size feedback is helpful, given the presence of feedback from rate mismatch. The feedback from routers to end-systems is time delayed, and may introduce instabilities and complex non-linear dynamics. Delay dynamical systems are often modeled using delay differential equations to facilitate a mathematical analysis of their performance and dynamics. The RCP models with and without queue size feedback give rise to two distinct non-linear delay differential equations. Earlier work on this design question was based on methods of linear systems theory. For further progress it is quite natural to employ nonlinear techniques. In this study, we approach this design question using tools from control and bifurcation theory. The analytical results reveal that the removal of queue feedback could enhance both stability and convergence properties. Further, using Poincar\'{e} normal forms and center manifold theory, we investigate two nonlinear properties, namely, the type of Hopf bifurcation and the asymptotic stability of the bifurcating limit cycles. We show that the presence of queue feedback in the RCP can lead to a sub-critical Hopf bifurcation, which would give rise either to the onset of large amplitude limit cycles or to unstable limit cycles. Whereas, in the absence of queue feedback, the Hopf bifurcation is always super-critical and the bifurcating limit cycles are stable. The analysis is complemented with computations and some packet-level simulations as well. In terms of design, our study suggests that the presence of both forms of feedback may be detrimental to the performance of RCP.
% In fact, ours is the first study that presents evidence to suggest that fluid models for congestion control algorithms may undergo a sub-critical Hopf bifurcation.
% In particular, we proceeded to analyze the dynamics of these algorithms as conditions for stability are just violated
\end{abstract}

\end{frontmatter}

\section{Introduction}
Most queuing systems routinely share information regarding waiting times, or queue lengths, with customers. Such information can certainly influence the behavior of customers. If such feedback is not instantaneous, but is time-delayed, it can have a significant impact on the underlying system dynamics; for example, see \citep{allon2011,novitzkysiads2019,pender2018,sharma2001}. The presence of feedback delays makes the system infinite-dimensional, and may pose numerous theoretical and practical challenges. In general, the stability of a closed-loop system is sensitive to feedback delays, which normally necessitates a detailed stability analysis. Local stability analysis retains only the linear component and ignores all higher order terms of the nonlinear system before addressing the issue of stability. However, the feedback delays of a nonlinear dynamical system may result in various complex dynamics like bifurcation, chaos, etc. So, it looks appealing to have an analytical methodology which may allow us to capture the impact of some nonlinear terms while performing a Taylor expansion of the nonlinear system about its equilibrium. Local bifurcation theory is one such methodology \cite{hassard1981}. For example, see \citep{dubeycnsns2019,huangcnsns2018,khoshcsf2019,novitzkysiads2019} for some stability and bifurcation analysis of dynamical systems with feedback delays. Moreover, without an understanding of the dynamics of the system in the unstable regime, choosing an operating point close to the boundary of the stable region could be risky. A comprehensive understanding of local bifurcation phenomena may help yield insights into the behavior of the system in the unstable regime. This paper employs both linear systems theory and non-linear techniques to investigate how the feedback of queue size can impact the system behavior in the setting of congestion control protocols for the Internet. We consider protocols where end-systems use feedback, which is time-delayed, from routers to adjust their rates. There is a continued interest in analyzing the stability and dynamical properties of fluid models for Internet congestion control algorithms \citep{gentilecnsns2014,liu2011nlarwatcp,peicnsns2019,peiijbc2018,voice2009maxminrcp,zhangcnsns2015}. In this paper, our focus will be on a well-known congestion control transport protocol called the Rate Control Protocol (RCP) \citep{balakrishnan2007stability,dukkipatircpac,krbookchap,krv2009,lakshmikantha2008}.

% There is a growing literature related to the development of congestion control protocols that could rely on explicit feedback \citep{he2017,jose2016xcc,liu2016xcc,liu2012XCPHopf,mahdian2016xcc,ren2016xcc}.
Internet congestion control has been an active area of networking research for several decades~\citep{srikant2012}. The Transmission Control Protocol (TCP) is currently the most commonly used transport protocol, which handles congestion control in the Internet. The widespread use of high bandwidth-delay networks and wireless links create limitations on the performance of standard TCP. Currently, TCP uses packet loss and delay as signals of network congestion. But, delay and loss are damage to packets, and hence this implicit signaling mechanism affects the performance limits and system Quality of Service (QoS). This has led to the development of congestion control protocols that could employ explicit feedback. Some examples of explicit congestion control algorithms include the Rate Control Protocol (RCP) \citep{balakrishnan2007stability,dukkipatircpac,krbookchap,krv2009,lakshmikantha2008}, eXplicit Control Protocol (XCP) \citep{he2017,katabi2002,liu2012XCPHopf}, JetMax \citep{zhangjetmax} and MaxNet \citep{wydrowskimaxnet}. Explicit congestion control protocols can convey accurate information to end-systems, which are then able to better regulate their flow and congestion control mechanisms. In the class of explicit feedback algorithms, the Rate Control Protocol (RCP)  \citep{balakrishnan2007stability,dukkipatircpac,krv2009} has the potential to offer stable and fair network performance, along with low latency and high link utilization. One architecture that could benefit from the use of RCP is a host-centric one, i.e., IP-based networks (for example, see \citep{baretto2015rcp,sharma17} and \citep{sun2012rcp}).
% In \citep{baretto2015rcp}, RCP has been explored in wireless environments. In \citep{sharma17}, the feasibility of implementing RCP with flexible packet processing architectures has been demonstrated, especially in data center networks. Simulation results in \citep{sun2012rcp} reveal that RCP outperforms TCP in terms of throughput and queue size in satellite networks. 
Another architecture that appears to be appealing for RCP are future data-centric networks, which are called Named Data Networking~(NDN) \citep{zhangndn2010}. In NDN, there is no IP address, and all data are named with unique names.
% For example, a video produced by PARC may have the name /parc/videos/Widget1.mpg. To fetch any desired data, the user sends out an interest packet which carries the name that identifies the data.
Also, the data can be retrieved through multiple paths from multiple sources. In fact, the implicit feedback mechanism used by TCP is not reliable in these environments \citep{ren2016xcc}. Therefore, researchers tend to focus on employing rate-based RCP-style algorithms in NDN; for e.g., see \citep{lei15,mahdian2016xcc} and \citep{zhong2017}.

% In this paper, we address a key design problem that arises in the study of feedback mechanism used by RCP. To expand further, 
RCP aims to achieve processor sharing by assigning a single fair rate for all the flows traversing the bottleneck link. The dynamics and performance of a protocol are largely affected by the mechanisms that routers employ to control the congestion. Currently, regardless of networking architectures, RCP computes the fair rate using two forms of feedback: rate mismatch and the queue size. An outstanding question in the design of RCP is whether two forms of feedback are needed in the protocol specification. We address this question by evaluating the performance of RCP with and without two forms of feedback. Early results in~\citep{krv2009} show that the presence of queue feedback in RCP may lead to less accurate control over the queue size. However, this insight was based on some initial simulations, and more analysis would be required before arriving at a better understanding of this design consideration. An RCP router utilizes a field in the packet header to convey the fair share rate at which the flows can send data into the network. However, the feedback about the fair rate to end-systems is not instantaneous. Therefore, RCP works like a closed-loop control system with feedback delays. Hence, the stability becomes an important consideration in the design of RCP. Apart from ensuring stability, another important design objective is to make sure that the system converges quickly to a stable equilibrium. Therefore, we explore the impact of queue feedback on the local stability and the convergence properties of RCP. It is also natural to investigate the nature of instabilities that may occur if the stability conditions are just violated. We hasten to add that we are not interested in destabilizing the network, but wish to employ the tools offered by local instability analysis to gain some insights into the non-linear properties of the system under consideration. Thus, we also conduct a detailed local Hopf bifurcation analysis to investigate the effects of queue feedback on the non-linear dynamical properties of RCP. Taken together, we consider local stability, convergence rate, the type of Hopf bifurcation, and the stability of limit cycles as the performance metrics to deduce whether the queue feedback is beneficial or not.

In a control theoretic approach, congestion control algorithms are often modeled as delay differential equations. The systems with and without feedback based on queue size give rise to different nonlinear delay differential equations. For now, we consider the RCP model that assumes flows with the same round-trip time, operating over a single bottleneck link. Here, round-trip time (RTT) denotes the sum of all the delays from source to link, and from link to source.  Instability in the system could be induced by varying some of the system parameters. But, variations of some system parameters may affect the equilibrium values, which is undesirable. Moreover, to conduct a unified analysis of both the design choices, it is desirable to use a common exogenous parameter to push the system just into the unstable regime.
 
Previous studies of the stability of RCP were confined to developing sufficient conditions to ensure local stability~\citep{balakrishnan2007stability,krv2009}. In this work, we derive necessary and sufficient conditions to ensure local asymptotic stability of RCP for the cases with and without queue size feedback. This enables us to determine the stability region in the parameter space. It is then shown that, as the bifurcation parameter varies, the system where feedback is based on both rate mismatch and queue size, readily loses local stability through a Hopf bifurcation~\citep{hassard1981}. Similarly, the rate of convergence analysis reveals that the convergence rate to equilibrium increases when the queue size term is absent. The theoretical framework we employ to study the nature of the Hopf bifurcations is the 
Poincar\'{e} normal form \citep{hassard1981} which requires the computation of a center manifold existing near the degenerate equilibrium point under investigation and the determination of the flow on this manifold. Using this framework, we analytically characterize the type of
Hopf bifurcation, and the nature of the bifurcating limit cycles. In \citep{voice2009maxminrcp}, it was shown that the RCP which uses only rate mismatch feedback always exhibit a super-critical Hopf bifurcation, and the emerging limit cycles are asymptotically stable. So far, the bifurcation properties for the case where we have both rate mismatch and queue size feedback have not been studied. In this paper, we show that the RCP which uses both rate mismatch and queue size feedback can exhibit a sub-critical Hopf bifurcation, for some parameter values. A sub-critical Hopf bifurcation is undesirable for real engineering systems as a small perturbation around the system equilibrium may give rise to either limit cycles with large amplitude, or unstable limit cycles \citep{strogatz2018}. Thus, the queue feedback in the RCP model could create adverse effects on the stability of the limit cycles.

To the best of our knowledge, this is the \emph{first study} that presents strong evidence to suggest that fluid models representing Internet congestion control algorithms can undergo a sub-critical Hopf bifurcation, for some parameter values. Therefore, this study reveals some novel insights into the non-linear dynamical properties which has been overlooked in previous studies of congestion control algorithms. Moreover, in general, the insights from Hopf bifurcation analysis could guide design considerations such that any loss of local stability only occurs via the emergence of small amplitude stable limit cycles. In other words, the nature of Hopf bifurcation and the stability of the bifurcating limit cycles should also be considered while designing congestion control protocols. Hence, this work could serve as an important step towards better design guidelines of the Internet congestion control algorithms. In essence, all the analytical results of this study tend to favor the design choice that uses only rate mismatch feedback. For such a system, a necessary and sufficient condition for non-oscillatory convergence is derived. This condition helps in finding an optimal value for a key protocol parameter. Bifurcation diagrams, numerical computations and packet-level simulations serve to validate some of the theoretical insights.   
% In fact, in the context of congestion control algorithms, the possibility of occurrence of a sub-critical Hopf has not been extensively studied so far. To our knowledge, this is the first study the possibility of a sub-critical Hopf bifurcation in the congestion control algorithms has not been much investigated in detail. Therefore, this study revealed some novel insights into the dynamical properties of our insights signals the need for additional studies to understand more about this  
% 

The rest of this paper is structured as follows. In Section 2, we outline the non-linear fluid model for RCP. In Section 3, we investigate the local stability of RCP in the presence and absence of queue size feedback. The rate of convergence and local Hopf bifurcation analysis are outlined in Sections 4 and 5. A necessary and sufficient condition for non-oscillatory convergence is derived in Section 6. Finally, in Section 7, we summarize our key insights and suggest some avenues for further research.
%%%%%%%%%%%%%%%%%%%%%%%%%%%%%%%%%%%%%%%%%%%%%%%%%%%%%%%%%%%%%%%%%%%%%%%%%%%%%%%%

\section{RCP Model}
At the fluid level, modeling the congestion control algorithms using the framework of delay differential equations has enabled their analysis to be subjected to tools from control and bifurcation theory. This section describes the non-linear model that represents the dynamics of the RCP protocol. RCP calculates the fair rate for all flows sharing a single bottleneck link by using feedback based on rate mismatch and the instantaneous queue size. The model for RCP is governed by the following non-linear delay differential equation~\citep{dukkipatircpac,voice2009maxminrcp}
\begin{equation}
\label{eq:model_RCP}
\frac{d}{dt}R(t) = \dfrac{R(t)}{C\overline{T}}\left(a\big(C-y(t)\big)-\beta\dfrac{q(t)}{\overline{T}}\right),
\end{equation}
where
\begin{equation}
\label{eq:queue_equ}
\begin{aligned}
y(t) &= \sum\limits_{s}R(t-T_s),\\
\frac{d}{dt}q(t) &= \begin{cases} \begin{array}{l} \big[y(t)-C\big] \\  \big[y(t)-C\big]^+ \end{array} \begin{array}{l} q(t)>0\\
 q(t)=0, \end{array}
\end{cases}
\end{aligned}
\end{equation}
using the notation $z^+=\text{max}(z,0)$. Here $R(t)$ denotes the rate that
RCP updates for all flows passing through the link, $y(t)$ is the aggregate load arriving at the link, $C$ is the capacity of the link, $q(t)$ is the instantaneous queue size, $T_s$ is the round-trip time (RTT) experienced by the traffic flow $s$, $\overline{T}$ represents the average RTT of packets passing through the link, and $a$, $\beta$ are non-negative dimensionless protocol parameters.

The equation for the queue dynamics is \\
\begin{equation}
\label{eq:modq}
\begin{aligned}
\frac{d}{dt}q(t) & = \big[y(t) - C\big] \qquad \forall\  q\left(t\right).
\end{aligned}
\end{equation}
\newcommand{\myk}{\tilde{k}}
%*************** Local Stability  *************************
\section{Local stability analysis}
% The initial, and in fact very common, style of analysis for non-linear
% time delayed systems is to first linearize the equation and then study the
% stability properties of the linear system. 
The initial, and in fact very common, style of stability analysis for non-linear time delayed systems is to first linearize the equation and then study the stability properties of the linearized system. In this section, the non-linear fluid model of RCP is first linearized about its equilibrium, and then attractive conditions for local stability guide design recommendations.
% Since we analyze only the local stability, the linearization would give sufficient information to deduce whether or not the system converges to equilibrium. 
For the sake of simplicity, it is assumed that the bottleneck link carries flows with the same round-trip time $\tau$. It is preferable not to use any of the system parameters as the bifurcation parameter, as varying them would affect the system equilibrium. Instead, an exogenous non-dimensional parameter, $\kappa$, is used to drive the system just into the regime of local instability. This has various advantages. we need not be concerned with the dimension of the parameter, and as it is common for both the design choices we can compare the results fairly.
% It is then shown that as the bifurcation parameter varies, the system whose feedback is based on both rate mismatch and the instantaneous queue size, loses its stability through a Hopf bifurcation.
\subsection{Feedback based on rate mismatch and queue size}
The RCP model under consideration can now be represented as
\begin{equation}
\begin{aligned}
\frac{d}{dt}R(t) &= \dfrac{\kappa R(t)}{C\tau}\Bigg(a\big(C-R(t-\tau)\big)-\beta\dfrac{q(t)}{\tau}\Bigg),\label{eq:simpmodel_RCP}\\
\frac{d}{dt}q(t) &= \kappa\big(R(t-\tau)-C\big). %\label{eq:simpmodq}
\end{aligned}
\end{equation}
Let $(R^{*},q^{*})$ represents the non-trivial equilibrium of \eqref{eq:simpmodel_RCP}, then
\begin{equation}
\begin{aligned}
\label{eq:eqbmeqns}
R^{*} &= C,\\
q^{*} &= 0.
\end{aligned}
\end{equation}
From \eqref{eq:eqbmeqns}, it can be noted that the equilibrium values are independent of the bifurcation parameter $\kappa$. To linearize a non-linear system, we write the Taylor series expansion about the equilibrium point and include only the linear terms. Consider the perturbation $r(t)=R(t)-R^{*}$, and linearize \eqref{eq:simpmodel_RCP} about the equilibrium to obtain
\begin{equation}
\begin{aligned}
\label{eq:linear}
\frac{d}{dt}r(t) &= -\kappa\left(\frac{a}{\tau} r(t-\tau) +\frac{\beta}{\tau^{2}}q(t)\right),\\
\frac{d}{dt}q(t) &= \kappa r\left(t-\tau\right).
\end{aligned}
\end{equation}
The characteristic equation for the linearized system \eqref{eq:linear} is
\begin{equation}
\label{eq:charwithBeta}
\begin{aligned}
\lambda^{2}\tau^{2}e^{\lambda \tau} + a\kappa\tau\lambda + \kappa^{2}\beta = 0.
\end{aligned}
\end{equation}
By analyzing the roots of the above transcendental characteristic equation, we proceed to derive a necessary and sufficient condition for local asymptotic stability. For the system to be stable, all the roots of \eqref{eq:charwithBeta} should lie on the open left-half, $\mathbf{Re}(\lambda)<0$, of the complex plane. The system becomes locally unstable when the roots cross the imaginary axis and go to the right half of the complex plane. Therefore, the condition for the crossover defines the
bounds on the model parameters which guarantees local asymptotic stability. To find the critical value of $\kappa$, at which the characteristic equation \eqref{eq:charwithBeta} has a pair of purely imaginary roots, substitute $\lambda=i\omega_0,\ \omega_0 > 0$ in
\eqref{eq:charwithBeta}. Then, separating real and imaginary terms, and equating them to zero gives
\begin{equation}
\begin{aligned}
\label{eq:reim}
\  \  \ -(\omega_0\tau)^{2}\cos\left(\omega_0\tau\right) + \kappa^{2}\beta &= 0,
\end{aligned}
\end{equation}
\begin{equation}
\begin{aligned}
\label{eq:reim1}
-(\omega_0\tau)^{2}\sin\left(\omega_0\tau\right) + \kappa a\omega_0\tau &= 0.
\end{aligned}
\end{equation}
Solving \eqref{eq:reim} and \eqref{eq:reim1}, yields
\begin{equation}
\label{eq:eqlty}
\kappa = \frac{1}{\theta}\sin^{-1}\left(\frac{a}{\theta}\right),
\end{equation}
where 
\begin{equation}
\theta = \dfrac{\omega_0\tau}{\kappa} =\sqrt{\frac{a^{2} + \sqrt{a^{4} + 4\beta^{2}}}{2}}.
\end{equation}
Now, the critical value ($\kappa_c$) at which \eqref{eq:charwithBeta} has a purely imaginary root is given by 
%  \begin{equation}
% \label{eq:beta_c}
% \sqrt{\frac{a^{2} + \sqrt{a^{4} + 4{\beta_{c}}^{2}}}{2}} = \sin^{-1}\left(\frac{a}{\sqrt{\frac{a^{2} + \sqrt{a^{4} + 4{\beta_{c}}^{2}}}{2}}}\right).
% \end{equation}
  
% Now, we obtain the critical value of the bifurcation parameter, at which the characteristic equation \eqref{eq:charwithBeta} has a pair of roots on the imaginary axis as 
\begin{equation}
\label{eq:hopfcond}
\kappa_c = \frac{1}{\sqrt{\frac{a^{2} + \sqrt{a^{4} + 4{\beta}^{2}}}{2}}}  \sin^{-1}\left(\frac{a}{\sqrt{\frac{a^{2} + \sqrt{a^{4} + 4{\beta}^{2}}}{2}}}\right).
\end{equation}

To exhibit that the system loses local asymptotic stability via a Hopf bifurcation, as $\kappa$ increases beyond $\kappa_c$, the following transversality condition~\citep{hassard1981} has to be verified:
$$\mathbf{Re}\left(\dfrac{d\lambda}{d\kappa}\right)_{\kappa=\kappa_c}\neq 0.$$\\
Differentiating \eqref{eq:charwithBeta} with respect to $\kappa$ gives
$$\dfrac{d\lambda}{d\kappa} =  \dfrac{\lambda(a\lambda\tau + 2\kappa\beta)}{(2+\lambda\tau)(a\kappa\lambda\tau + \kappa^{2}\beta) -\kappa a\lambda \tau }.$$\\
It is clear that
\begin{equation*}
   \mathrm{Sign}\,\Bigg(  \mathbf{Re}\left( \frac{d\lambda}{d\kappa} \right) \Bigg)_{\kappa=\kappa_c} 
   = \mathrm{Sign}\,\left(  \mathbf{Re}\left(\left( \frac{d\lambda}{d\kappa} \right)^{-1} \right)\right)_{\kappa=\kappa_c}.	
\end{equation*}
Therefore, instead of finding ($d \lambda/d \kappa$), consider its inverse, i.e., $\left({d\lambda}/{d\kappa}\right)^{-1}$. Further simplification gives\\
\begin{equation}
\label{eq:transversalitycheck}
 \left(\dfrac{d\lambda}{d\kappa}\right)^{-1}= \kappa\left(\frac{1}{\lambda}+ \frac{\tau(a\lambda\tau+\kappa\beta)}{(a\lambda\tau+2\kappa\beta)}\right).
\end{equation}
Evaluating the real part of \eqref{eq:transversalitycheck} at $\kappa=\kappa_c$ yields\\
\begin{equation*}
   \Bigg(\mathbf{Re}\left( \frac{d\lambda}{d\kappa} \right)^{-1} \Bigg)_{\kappa=\kappa_c} 
   =   \kappa_c\tau\bigg( \frac{a^2\omega_0^2\tau^2 + 2\kappa_c^2\beta^2}{a^2\omega_0^2\tau^2 + 4\kappa_c^2\beta^2} \bigg)  > 0,
\end{equation*}
which satisfies the transversality condition.
% from which, we obtain
% $$\mathbf{Re}\left(\dfrac{d\lambda}{d\kappa}\right)_{\kappa=\kappa_c}=\dfrac{\omega_0^{2}\tau\left(a^{2}\omega_0^{2}\tau^{2}\kappa_c+2\beta^{2}\kappa_c^{3}\right)}{A^2+B^2} > 0,$$
% where 
% \begin{eqnarray}
% % \omega_0 &=& \pi/(\tau_1+\tau_2),\nonumber\\
% \begin{aligned}
% \quad \quad \quad \quad A&=2\kappa_c^{2}\beta-a\kappa_c(\omega_0\tau)^{2},\nonumber\\
% \quad \quad \quad \quad B&=\omega_0\tau\kappa_c^{2}\beta + a\kappa_c\omega_0\tau.\nonumber
% \end{aligned}
% \end{eqnarray}
The positivity of the above derivative implies that the system transits from stability to instability through a Hopf bifurcation, at $\kappa=\kappa_c$. It also means that the roots of \eqref{eq:charwithBeta} crossover the imaginary axis from left to right with positive velocity, and hence the system does not regain the stability with a further increase in $\kappa$. The linearized system \eqref{eq:linear} is asymptotically stable for all values of $\kappa$ less than $\kappa_c$, and unstable for $\kappa>\kappa_{c}$. Now, the necessary and sufficient condition for local stability of \eqref{eq:simpmodel_RCP} is
\begin{equation}
\label{eq:nscond}
\kappa{\theta} < \sin^{-1}\left(\frac{a}{\theta}\right).
\end{equation}
By setting $\kappa$ = 1 (to get back to the original system), \eqref{eq:nscond} can be rewritten as
\begin{equation}
\label{eq:nscond1}
\sqrt{\frac{a^{2} + \sqrt{a^{4} + 4\beta^{2}}}{2}} < \sin^{-1}\left(\frac{a}{\sqrt{\frac{a^{2} + \sqrt{a^{4} + 4\beta^{2}}}{2}}}\right).
\end{equation}
From \eqref{eq:nscond1}, it can be observed that the protocol parameters $a$ and $\beta$ play a vital role in ensuring system local stability. Figure \ref{fig:schart} graphically represents the Hopf condition and the region of local stability of RCP. This makes it easier to understand the relationship between $a$ and $\beta$ to ensure stability.
\newcommand{\wdth}{\textwidth}
\begin{figure}[hbtp!]
 \centering
 \psfrag{B}{{\hspace{-0.8cm} \small Parameter, $\beta$}}
 \psfrag{A}{{\hspace{-0.8cm}\small Parameter, $a$}}
 \psfrag{stable}{\footnotesize stable region}
 \psfrag{unstable}{\small }
 \psfrag{Hopf}{\footnotesize Hopf condition}
 \psfrag{0}{\small{$0$}}
% \psfrag{4}{$\small{4}$}
% \psfrag{2}{$\small{2}$}
\psfrag{0}{\hspace{0.3cm}\small{$0$}}
\psfrag{0.2}[][]{\small{$0.2$}}
\psfrag{0.4}{\small{$0.4$}}
\psfrag{0.6}{\small{$0.6$}}
\psfrag{0.0}{\hspace{0.1cm}\small{$0$}}
\psfrag{0.00}{\hspace{0.2cm}\small{$0$}}
\psfrag{0.79}[][]{\small{$\pi/4$}}
\psfrag{1.57}{\small{$\pi/2$}}
 \includegraphics[width=0.7\wdth]{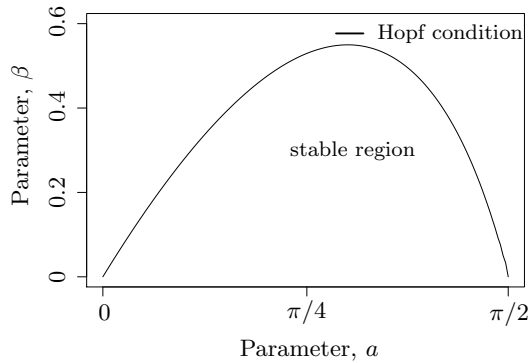}
 \caption{Local stability chart for \eqref{eq:simpmodel_RCP}, highlighting the relationship between the protocol parameters $a$ and $\beta$ to ensure local stability.}
 \label{fig:schart}
 \end{figure}
%This also enables us to identify the bounds on the protocol parameters to ensure local stability.  
 % In the theoretical analysis done so far, we exhibited that the system loses local stability via a Hopf bifurcation, as the bifurcation parameter crosses a critical value.  
\subsection{Feedback based on only rate mismatch}
\begin{figure}[hbtp!]
\centering
\psfrag{R}{\hspace{-0.28cm}  \small Rate}
\psfrag{a}{{\hspace{-1.5cm} \small Bifurcation parameter, $\kappa$}}
\psfrag{0.95}{\hspace{0cm}\small{$0.95$}}
\psfrag{1.00}{\hspace{0cm}\small{$1.00$}}
\psfrag{1.05}{\hspace{0cm}\small{$1.05$}}
\psfrag{0.0}{\small{$0.0$}}
\psfrag{0}{\small{$0$}}
\psfrag{0.5}{\hspace{0cm}\small{$0.5$}}
\psfrag{1.0}{\small{$1.0$}}
\psfrag{2.0}{\small{$2.0$}}
\psfrag{1}{\small{$1$}}
\psfrag{1.5000}{\small{$1.5$}}
\psfrag{2}[][]{\small{$2$}}
\psfrag{1.5707}{\small{$\pi/2$}}
\includegraphics[width=0.55\wdth]{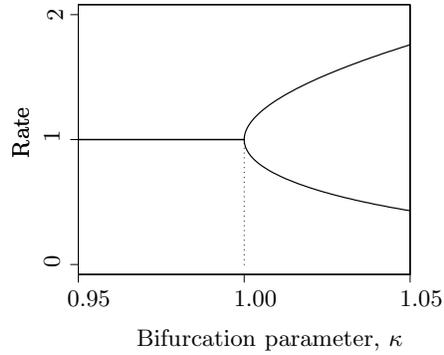}
\caption{Bifurcation diagram showing the emergence of limit cycles as the bifurcation parameter $\kappa$ increases beyond $1$. The parameter values used are $a = 1.5$, $\beta =0.1$, $C=1$ and $\tau=1$. For theses parameter values, the critical value at which Hopf occurs is $\kappa=1$.}
\label{fig:bfd_withq}
\end{figure}

In this subsection, local stability analysis is performed for RCP which uses only rate mismatch feedback. To remove the queue term from the RCP model, $\beta$ is set to zero in \eqref{eq:simpmodel_RCP}. This gives
\begin{equation}
\frac{d}{dt}R(t) = \dfrac{\kappa R(t)}{C\tau}\Bigg(a\big(C-R(t-\tau)\big)\Bigg)\label{eq:RCPmodel_withnoq}. %\label{eq:simpmodq}
\end{equation}
Linearizing \eqref{eq:RCPmodel_withnoq} about its equilibrium gives 
\begin{equation}
\label{eq:linear_withnoq}
\frac{d}{dt}r(t) = -\kappa\left(\frac{a}{\tau}r(t-\tau)\right).
\end{equation}
The characteristic equation of \eqref{eq:linear_withnoq} can be written as
\begin{equation}
\label{eq:chareq_withnoq}
\lambda + \left(\frac{\kappa a}{\tau}\right)e^{-\lambda \tau} = 0.
\end{equation}
Proceeding as outlined in the previous subsection, the necessary and sufficient condition for local asymptotic stability of \eqref{eq:RCPmodel_withnoq} can be written as
\begin{equation}
\label{eq:hopfcond_withnoq}
 a \kappa < \frac{\pi}{2}.
\end{equation}
Substituting $\kappa$ = 1 gives 
\begin{equation}
\label{eq:nscond1_withnoq}
 a < \frac{\pi}{2}.
\end{equation}
\noindent \emph{Impact of queue size feedback on stability}: As compared to \eqref{eq:nscond1}, the RCP model which uses only rate mismatch feedback gives a simple stability condition, which makes it easier to design and ensure a stable system. It should be noted that condition \eqref{eq:nscond1_withnoq} is necessary but not sufficient for the system \eqref{eq:simpmodel_RCP} to be locally stable. Now, if the value of protocol parameter $a \in(0,\frac{\pi}{2})$, then the RCP system with no queue size feedback is locally stable, whereas the system with queue feedback loses its local stability through a Hopf bifurcation, as parameter $b$ varies beyond some threshold. The Hopf bifurcation would lead to the emergence of limit cycles in the system dynamics. These limit cycles can manifest themselves in the queue size, which could degrade performance.
% makes it difficult to control the queuing delay. This behavior could also result in bursty packet loss, jitter (delay variations), which would severely affect the performance of delay-sensitive applications. 
Hence, the presence of queue size feedback seems to be detrimental to the system stability.
A numerical example is now given to illustrate the existence of Hopf bifurcation and the emergence of limit cycles.
%%%%%%%%%%%%%%%%%%%%%%%%%%%%%%%%%%%PHASE PORTRAITS%%%%%%%%%%%%%%%%%%%%%%%%%%%
\begin{figure}[hbtp!]
\centering
\psfrag{x}{\hspace{-0.3cm}  \small  $R(t)$}
\psfrag{y}{\hspace{-0.6cm}\vspace{5mm}  \small$R(t-\tau)$}
\psfrag{eqbm}{\hspace{-0.5mm}\footnotesize{Equilibrium}}
\psfrag{231}{\hspace{0.1cm}\small{$0$}}
\psfrag{236}{\hspace{-0.1cm}\small{$5000$}}
\psfrag{241}{\hspace{-0.2cm}\small{$10000$}}
\psfrag{2.0}{\footnotesize{$2.0$}}
\psfrag{0}{\hspace{-0.0cm}\footnotesize{$0$}}
\psfrag{0.5}{\hspace{-0.0cm}\footnotesize{$0.5$}}
\psfrag{1.0}{\hspace{0.0cm}\footnotesize{$1.0$}}
\psfrag{1.5}{\hspace{-0.0cm}\footnotesize{$1.5$}}
\psfrag{0.0}{\hspace{-0cm}\footnotesize{$0.0$}}
\psfrag{60}{\hspace{-0.07cm}\small{$60$}}
\psfrag{120}{\hspace{-0.07cm}\small{$120$}}
\begin{tabular}{c}
\subfloat[ $\kappa=0.95$]{\includegraphics[width=0.55\wdth]{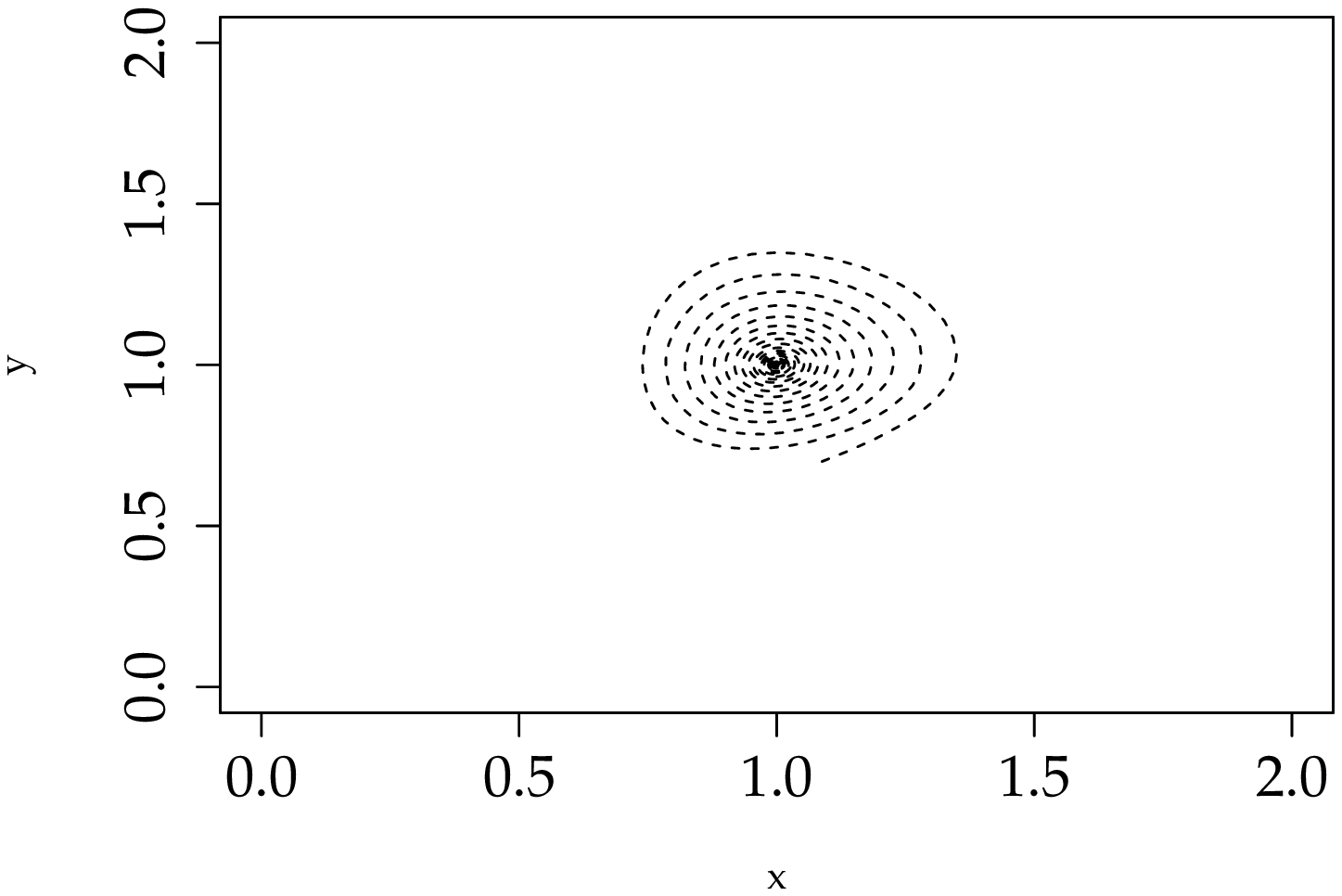}}\hspace{-9mm}
\subfloat[ $\kappa=1.05$]{\includegraphics[width=0.55\wdth]{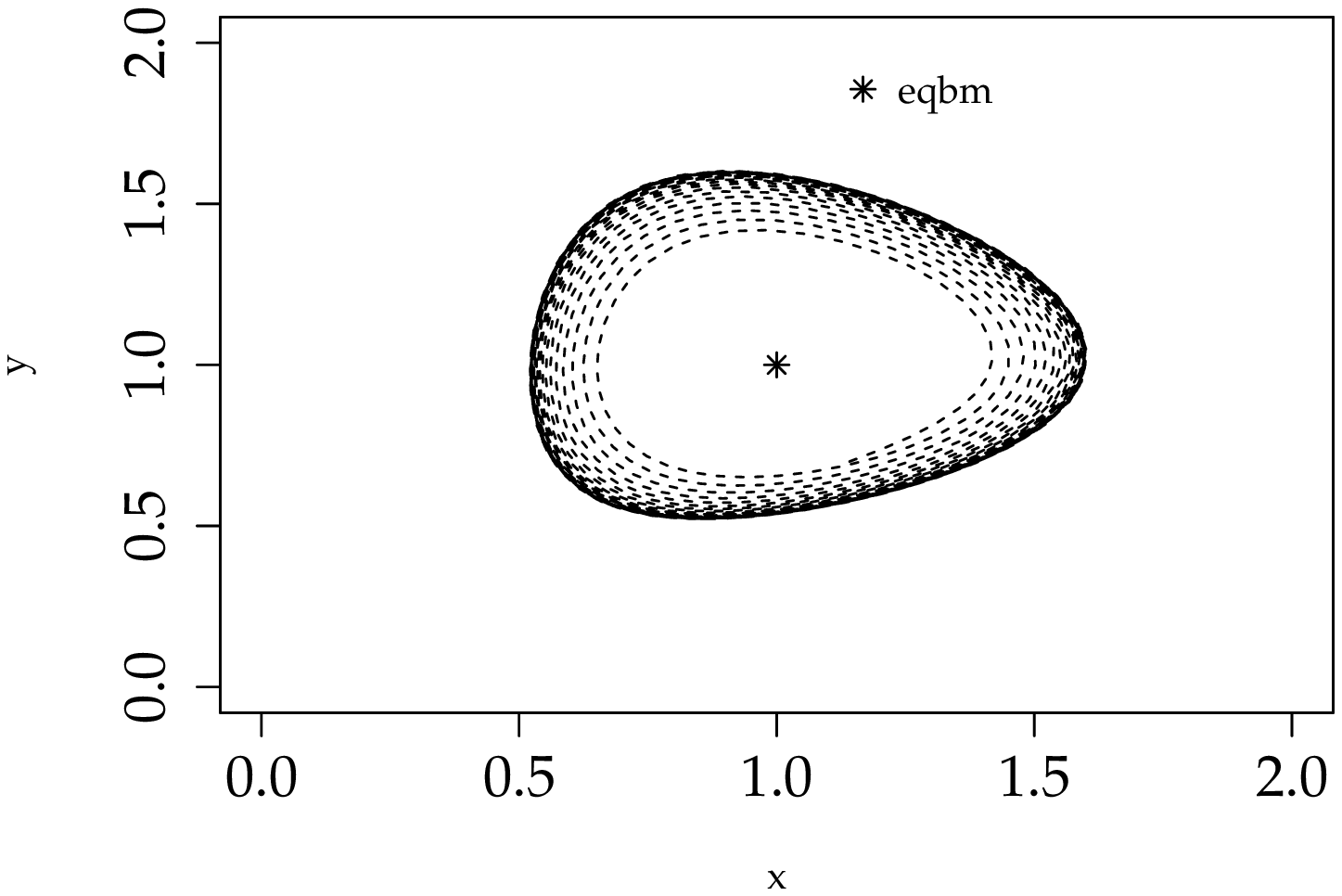}} 
\end{tabular}
\caption{Phase portraits of equation \eqref{eq:simpmodel_RCP} highlighting (a) convergence of the rate to a stable equilibrium, $R^{*}=1$ for $\kappa=0.95$, and (b) existence of a stable limit cycle for $\kappa=1.05$. As $\kappa$ increases beyond the critical value, the system dynamics exhibits a qualitative change from a stable
fixed point to a limit cycle. The values of the parameters used are $a = 1.5$, $\beta =0.1$, $C=1$ and $\tau=1$.}
\label{fig:phaseportrait}
 \end{figure}
%\vspace{0.5mm}\\
\\
\emph{Numerical Example}: Let $a = 1.5$, $\beta=0.1$, $C=1$ and $\tau=1$. For these parameter values, using the Hopf condition \eqref{eq:hopfcond}, the critical value of the bifurcation parameter that drives the system to the edge of the stable regime is $\kappa_c=1$. If $\kappa$ goes beyond this critical value, the system will lose its local stability via a Hopf bifurcation which leads to the emergence of limit cycles. To verify this, the phase portraits of \eqref{eq:simpmodel_RCP} for the cases: $\kappa < \kappa_c$ and $\kappa > \kappa_c$ are plotted in Figure \ref{fig:phaseportrait}. Using \eqref{eq:eqbmeqns}, we obtain the non-zero equilibrium of the system with the above choice of parameter values as $R^*=1$.
% We first start by choosing a value of the bifurcation parameter such that the system is locally asymptotically stable.
Based on the analytical results, for the value of $\kappa=0.95 < \kappa_c$, the system should be locally stable. Indeed, as shown in Figure \ref{fig:phaseportrait} (a), it can be observed that the rate $R(t)$ converges to the stable equilibrium ($R^*$), which implies that the system is locally asymptotically stable. We now marginally increase $\kappa$ beyond $\kappa_c$ (set $\kappa=1.05$), thus pushing the system into the unstable region. As expected, the system exhibits limit cycles (see Figure \ref{fig:phaseportrait} (b)). The numerical simulations were done using the numerical computing software MATLAB.
The bifurcation diagram shown in Figure \ref{fig:bfd_withq} was drawn using the MATLAB package DDE-Biftool~\citep{ddetool1,ddetool2}. Figure \ref{fig:bfd_withq} shows the existence of a super-critical Hopf as the bifurcation parameter $\kappa$ increases beyond $\kappa_c$.
%%%%%%%%%%%%%%%%%%%%%%%%%%%%%
%%%%Bifurcation Diagram
%%%%%%%%%%%%%%%%%%%%%%%%%%%%%

\begin{figure}[hbtp!]
\centering
\psfrag{t}{\hspace{-0.6cm} \vspace{-1.2cm} \footnotesize Time, $t$}
\psfrag{q}{\hspace{-1.3cm}\vspace{-1.3cm}  \footnotesize Queue size [pkts]}
\psfrag{a}{\hspace{-1.8cm} \vspace{-2.5cm}  \footnotesize (a) With $q(\cdot)$ feedback}
\psfrag{b}{\hspace{-1.8cm} \vspace{-2.5cm} \footnotesize (b) Without $q(\cdot)$ feedback}
\psfrag{14}{\hspace{-0.2cm}\footnotesize{$4000$}}
\psfrag{18}{\hspace{-0.2cm}\footnotesize{$8000$}}
\psfrag{0}{\footnotesize{$0$}}
\psfrag{60}{\hspace{-0.07cm}\footnotesize{$60$}}
\psfrag{120}{\hspace{-0.07cm}\footnotesize{$120$}}
\psfrag{10}{\hspace{0cm}\vspace{-1.5cm}\footnotesize{$0$}}
\psfrag{37}{\hspace{0cm}\footnotesize{$0$}}
\psfrag{38}{\hspace{0cm}\footnotesize{$0$}}
\psfrag{15}{\hspace{-0.2cm}\footnotesize{$5000$}}
\psfrag{41}{\hspace{-0.2cm}\footnotesize{$4000$}}
\psfrag{43}{\hspace{-0.2cm}\footnotesize{$5000$}}
\psfrag{45}{\hspace{-0.35cm}\footnotesize{$8000$}}
\psfrag{48}{\hspace{-0.35cm}\footnotesize{$10000$}}
\psfrag{20}{\hspace{-0.35cm}\footnotesize{$10000$}}
% \begin{tabular}{c}
% \subfloat[ $a = 0.5,\   \beta = 0$]{\includegraphics[width=0.35\wdth, angle=270]{queue_size_stablenew.eps}}\hspace{1mm}%prev width = 2.15 in
% \subfloat[ $a = 0.5,\   \beta = 1$]{\includegraphics[width=0.35\wdth, angle=270]{queue_size_unstablenew.eps}} 
% \end{tabular}
%\begin{tabular}{c}
\includegraphics[width=0.35\wdth, angle=270]{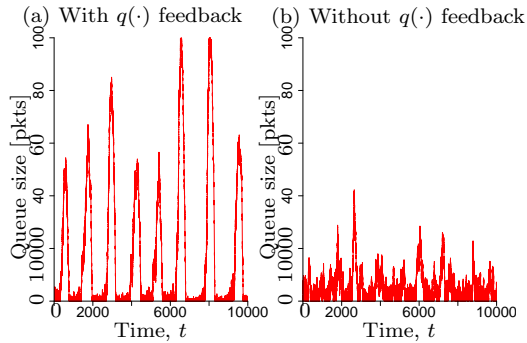}\hspace{1mm}%prev width = 2.15 in
%\end{tabular}
\caption{Traces from a packet-level simulation showing the evolution of queue size in a single
 bottleneck link with $C=1$ packet per unit time, $\tau=100$ time units and $100$ RCP sources. Observe that the queue appears to be stable for the choice of parameter values $a = 0.5$ and $\beta = 0$, whereas it begins to oscillate for $a = 0.5$ and $\beta = 1$.}
 \label{fig:simulations}
 \end{figure}

%In the next section, packet-level simulations will be presented to illustrate the effect of queue size feedback on the system stability.
%\section{Packet-level simulations}In this section,
Another essential compliment to the theoretical analysis is the packet-level simulations. The theoretical insights should be validated by investigating if the packet-level simulations of the underlying system exhibits the qualitative properties predicted through the analysis of the fluid model. The packet-level simulations are done using a discrete event RCP simulator (for more details, refer to \citep{krv2009}). The simulated network has a single bottleneck link setup that considers $C=1$ packet per unit time, number of flows = $100$ and RTT of all the flows as $\tau=100$ time units. Simulation traces in Figure \ref{fig:simulations} (a) and Figure \ref{fig:simulations} (b) show the evolution of queue size for the choice of parameter values $a = 0.5$, $\beta = 1$ and $a = 0.5$, $\beta = 0$, respectively. From \eqref{eq:nscond1_withnoq}, for the choice $a = 0.5$, $\beta = 0$, the system is expected to be stable. Indeed, this is confirmed in the simulation traces in Figure \ref{fig:simulations} (a) which does not exhibit any limit cycles in the queue size. Figure \ref{fig:simulations} (b) shows the emergence of limit cycles in the queue size. This is as expected, since the parameter values $a = 0.5$ and $\beta = 1$ violates the stability condition \eqref{eq:nscond1}, and lies outside the stable region (see Figure \ref{fig:schart}).
%So, these observations corroborate the occurrence of limit cycles due to the presence of queue size feedback.
%%%%%%%%%%%%%%%%%%%%%%%%%%%%PACKET LEVEL SIMULations

%%%%%%%%%%%%%%%%%%%%%%%%%%%%%%%%%%%%%%%%%%%%%%%%%%%%%%%%%%%%%%%%%%%%%%%%%%%%%%%%%%
\section{Rate of convergence}
% Local stability analysis revealed the nature of instabilities that may occur on violating those conditions. Now, within the stable regime, it is important to study how fast the system converges to equilibrium when perturbed. Especially, in the real network environments, these perturbations are more frequent due to the random arrival and departure of flows.
We now consider some convergence properties of the system in the stable regime. Rate of convergence is a key performance metric which must be considered for a congestion control algorithm. To that end, the impact of queue feedback on the convergence rate is examined by conducting the rate of convergence analysis in the presence and absence of queue size feedback.
\subsection{Feedback based on rate mismatch and queue size}
\begin{figure}[hbtp!]
 \centering
\psfrag{R}{\hspace{-1.8cm} \small{Rate of convergence, $\sigma$}}
\psfrag{a}{{\hspace{-0.8cm}\small{Parameter, $a$}}}
\psfrag{b=0}{\hspace{-2.1mm}\small{$\beta=0$}}
\psfrag{b=0.2}{\hspace{-2.1mm}\small{$\beta=0.2$}}
\psfrag{b=0.4}{\hspace{-2.1mm}\small{$\beta=0.4$}}
\psfrag{b=0.5}{\hspace{-2.1mm}\small{$\beta=0.5$}}
\psfrag{unstable}{\small }
\psfrag{roc}{\small Hopf condition}
\psfrag{0}{\small{$0$}}
\psfrag{0.0000000}{\hspace{0.4cm}\small{$0$}}
\psfrag{0.3678794}[][]{\small{$1/e$}}
\psfrag{0.7853982}{\hspace{0.3cm}\small{$\pi/4$}}
\psfrag{1.5707963}{\hspace{0.3cm}\small{$\pi/2$}}
\psfrag{0.0}{\hspace{0.0cm}\small{$0$}}
\psfrag{0.5}{\hspace{0.0cm}\small{$0.5$}}
\psfrag{1.0}[][]{\small{$1$}}
\psfrag{1.57}{\small{$\pi/2$}}
\includegraphics[width=0.75\wdth]{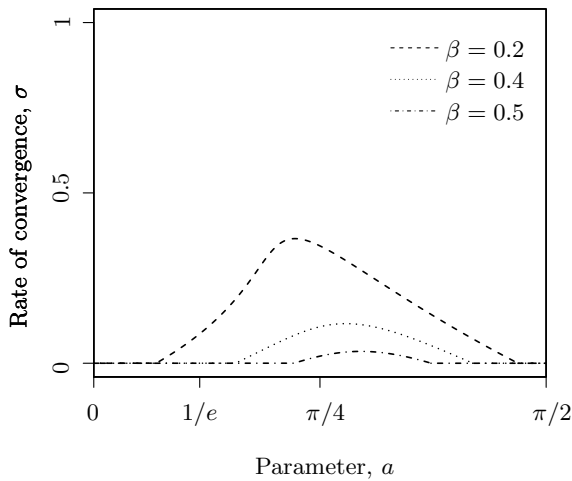}
\caption{Convergence rate to equilibrium of RCP for various values of the protocol parameter $\beta$. It can be observed that the rate of convergence decreases as the value of $\beta$ increases. The parameter values used are $C=10$ and $\tau=1$.}
\label{fig:roc_q2}
\end{figure}

To recapitulate, the characteristic equation of the linearized model of RCP which uses both rate mismatch and queue size feedback is
\begin{equation}
\label{eq:charwithBeta1}
\begin{aligned}
\lambda^{2}\tau^{2}e^{\lambda \tau} + a\tau\lambda + \beta = 0.
\end{aligned}
\end{equation}
From \eqref{eq:charwithBeta1} it can be observed that the characteristic function corresponding to the system is a second order quasi-polynomial. Therefore, in this case, it is difficult to get closed-form analytical expressions for the rate of convergence. In such a case, we resort to some numerical computation tools that can assist in finding the rightmost root of the characteristic equation. The real part of the rightmost root can give insights about the system behavior, and it can be used to calculate the rate of convergence. We use DDE-Biftool to analyze the rate of convergence for the system which includes queue feedback. For delay differential equations with real coefficients, the rightmost root could be a single real root or a complex conjugate pair. Figure \ref{fig:roc_q2} shows the rate of convergence computed numerically using DDE-Biftool, for various values of the protocol parameters. From Figure \ref{fig:roc_q2}, we can observe that the rate of convergence decreases as the value of $\beta$ increases. The parameter values chosen are $C=10$ and $\tau=1$.

\subsection{Feedback based on only rate mismatch}
In this section, following the style of analysis outlined in \citep{brauer79}, the rate of convergence analysis is performed for the RCP which uses only rate mismatch feedback. The analytical results enable us to investigate the impact of protocol parameters on the convergence rate. Here, we consider $\kappa=1$, to get back the original system. Now the characteristic equation can be written as 
\begin{equation}
\label{eq:ce_withoutq}
\lambda + \left(\frac{a}{\tau}\right)e^{-\lambda \tau} = 0.
\end{equation}
Substituting $\lambda\tau=x-\sigma\tau$ in \eqref{eq:ce_withoutq} yields
\begin{equation}\label{eq:newce_withoutq}
 (\sigma\tau-x)e^{x}-ae^{\sigma\tau}=0.
\end{equation} 
Here, $\sigma$ is considered to be the supremum of the solutions of \eqref{eq:newce_withoutq} over $(0,\infty)$ which guarantees that all the characteristic roots lie on the open left half of the complex plane. Let $-\alpha<0$ be the largest real part of all the roots of \eqref{eq:newce_withoutq}. Then, the rate at which the system approaches a stable equilibrium is given by $\sigma=(\alpha/\tau)$. A necessary and sufficient condition for all the eigenvalues of \eqref{eq:newce_withoutq} to lie on the left half-plane is \citep{hayes50}
\begin{align}
\sigma\tau&<1, \label{eq:nscroc1} \\
\sigma\tau&<ae^{\sigma\tau},\label{eq:nscroc2} \\
ae^{\sigma\tau}&<\frac{u}{\sin(u)}, \label{eq:nscroc3}
\end{align}
where $u$ is the solution of the equation
\begin{equation}
\label{eq:uroc}
 u=\sigma\tau \tan(u),
\end{equation}
in $0<u<\pi$, with $u=\pi/2$ if $\sigma=0$. Consider the following function
\begin{equation}
\label{eq:gofu}
 g(u)=\dfrac{u}{\sin(u)}e^{-u/\tan(u)},
\end{equation}
 which increases monotonically in the interval $u\in (0,\pi)$, with $g(0)=1/e$, $g(\pi/2)=\pi/2$ and $\lim_{u \rightarrow \pi}\ g(u)=\infty$. Now, using \eqref{eq:uroc} and \eqref{eq:gofu}, the inequality \eqref{eq:nscroc3} can be rewritten as
 \begin{equation}
  \label{eq:nscro3_1}
  a<g(u).
 \end{equation}
As $\sigma$ increases, $u$ decreases, and hence $g(u)$ is decreasing function of $\sigma$. Therefore, the maximum value of $\sigma$ that satisfies \eqref{eq:nscro3_1} can be obtained by solving its corresponding equality. By a similar argument, the left hand side of \eqref{eq:nscroc1} and \eqref{eq:nscroc2} increases with increase in $\sigma$. Thus, the maximum value of $\sigma$ that satisfies the inequalities \eqref{eq:nscroc1}, \eqref{eq:nscroc2} can be determined by solving the corresponding equalities. If the solution does not exist for any of these equations, then there is no restriction on the value of $\sigma$. Now, we summarize the results as follows.\\
Let $\sigma_1$, $\sigma_2$, $\sigma_3$ be the solutions of
\begin{align}
 \sigma \tau&=1,\label{eq:eqlty1} \\
 \sigma \tau e^{-\sigma\tau}&=a, \label{eq:eqlty2}\\
 u&=\sigma\tau \tan(u),\ \  g(u)=a, \label{eq:eqlty3} 
 \end{align}
respectively. Consider $\sigma_i=\infty,$ for $i=1,\ 2,\ 3$ if there is no solution exists for the corresponding equality. Then, the convergence rate $\sigma$ is given by
\begin{equation}
 \sigma=\min[\sigma_1,\sigma_2,\sigma_3].
\end{equation}
Now, the next step is to analyze the dependence of convergence rate on protocol parameter $a$, for $\tau>0$. The function $\sigma\tau e^{-\sigma\tau}$ reaches its maximum value of $1/e$ at $\sigma\tau=1$. Similarly the function $g(u)$ has a minima of $1/e$ at $u=0$. Let $a^{*}=1/e$, then there is no solution for \eqref{eq:eqlty2} if $a>a^{*}$, and for \eqref{eq:eqlty3} if $a<a^{*}$. Let $\sigma_2$ be the solution of \eqref{eq:eqlty2} on $0< a\leq a^{*}$. Similarly, consider $\sigma_3$ be the solution of \eqref{eq:eqlty3} on $a\ >\ a^{*}$. At $a=0$, it is obvious that the rate of convergence $\sigma=0$.
\paragraph{Case 1\ :\ $a \in (0,a^*)$} Differentiating \eqref{eq:eqlty2} with respect to $a$ gives
\begin{equation}
 \frac{d\sigma}{da}=\frac{e^{\sigma\tau}}{\tau(1-a e^{\sigma\tau})} \label{eq:sigmadash}.
\end{equation}
Using \eqref{eq:eqlty2}, the derivative \eqref{eq:sigmadash} can be written as 
\begin{equation}
 \frac{d\sigma}{da}=\frac{e^{\sigma\tau}}{\tau(1-\sigma\tau)}. \label{eq:sigmadash1}
\end{equation}
From \eqref{eq:sigmadash1}, it can deduced that $d\sigma_2 / da$ $>0$ if $\sigma_2\tau<1$. Hence $\sigma_2<\sigma_1$ for $a\in (0,a^*)$.
\paragraph{Case 2\ :\ $a=a^*$} Substituting $a=a^*=1/e$ in \eqref{eq:eqlty2} yields
\begin{equation}
 \sigma_2 \tau e^{-\sigma_2\tau}=a=1/e.
\end{equation}
It is known that the function $\sigma_2\tau e^{-\sigma_2\tau}$ reaches maximum of $1/e$ at $\sigma_2\tau=1$, thus $\sigma_2=\sigma_1=1/\tau$ at $a=a^*$.
\paragraph{Case 3\ :\ $a>a^*$} For $a\ >\ a^*$, using \eqref{eq:eqlty3}, the following holds
\begin{equation}
 g(u)=\dfrac{u}{\sin(u)}e^{-u/\tan(u)}>1/e. \label{eq:case4ineqlty}
\end{equation}
For $u \in (0,\pi)$ and $u/\sin(u) > 1$, \eqref{eq:case4ineqlty} can be written as 
\begin{equation}
 e^{-u/\tan(u)}>1/e \label{case4ineqlty1}.
\end{equation}
From \eqref{case4ineqlty1}, it can be deduced that $u/\tan(u)<1$, and hence $\sigma_3\tau<1$, and $\sigma_3<\sigma_1$.

\begin{figure}[hbtp!]
\centering
\psfrag{R}{\hspace{-1.8cm} \small Rate of convergence, $\sigma$}
\psfrag{a}{{\hspace{-0.8cm} \small Parameter, $a$}}
\psfrag{tau=1}{\hspace{-2mm}\small{$\tau=1$}}
\psfrag{tau=2}{\hspace{-2mm}\small{$\tau=2$}}
\psfrag{unstable}{\small }
\psfrag{roc}{\small Hopf condition}
\psfrag{0}{\small{$0$}}
\psfrag{0.0000000}{\hspace{0.4cm}\small{$0$}}
\psfrag{0.3678794}[][]{\small{$1/e$}}
\psfrag{0.7853982}{\hspace{0.3cm}\small{$\pi/4$}}
\psfrag{1.5707963}{\hspace{0.3cm}\small{$\pi/2$}}
\psfrag{0.0}{\hspace{0.0cm}\small{$0$}}
\psfrag{0.5}{\hspace{0.0cm}\small{$0.5$}}
\psfrag{1.0}[][]{\small{$1$}}
\psfrag{1.57}{\small{$\pi/2$}}
\includegraphics[width=0.7\wdth]{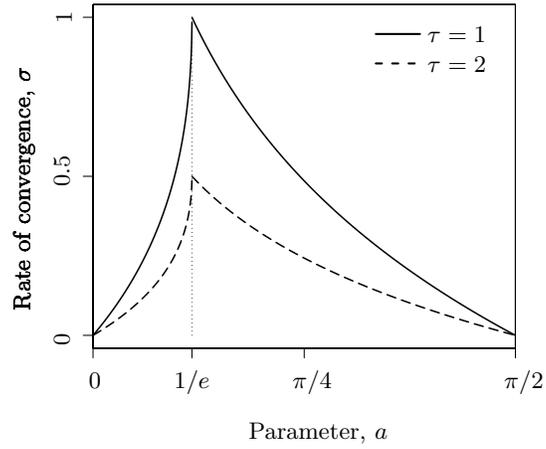}
\caption{Rate of convergence to equilibrium for RCP without queue feedback. The rate of convergence increases with $a$ and reaches maxima of $1/\tau$ at $a=1/e$, and then decreases for $a>1/e$.}
\label{fig:roc}
\end{figure}

\begin{figure}[hbtp!]
 \centering
\psfrag{b}{\hspace{-1.8cm} rate of convergence $\sigma$}
\psfrag{a}{{\hspace{-0.8cm}parameter $a$}}
\psfrag{tau=1}{\hspace{-2mm}\small{$\tau=1$}}
\psfrag{tau=2}{\hspace{-2mm}\small{$\tau=2$}}
\psfrag{0}{\small{$0$}}
\psfrag{time}{\hspace{-0.0cm}  \small Time}
\psfrag{Rate}{\hspace{-0.5cm}  \small Rate, $R(t)$}
\psfrag{0}{\small{$0$}}
\psfrag{5}{\hspace{0cm}\small{$5$}}
\psfrag{10}{\hspace{0cm}\small{$10$}}
\psfrag{15}{\hspace{0cm}\small{$15$}}
\psfrag{20}{\small{$20$}}
\psfrag{0}{\small{$0$}}
\psfrag{25}{\hspace{0cm}\small{$25$}}
\psfrag{50}{\small{$50$}}
\psfrag{0.0}{\hspace{0.0cm}\small{$0$}}
\psfrag{0.5}{\hspace{0.0cm}\small{$0.5$}}
\psfrag{1.0}[][]{\small{$1$}}
\psfrag{1.57}{\small{$\pi/2$}}
\includegraphics[width=0.7\wdth]{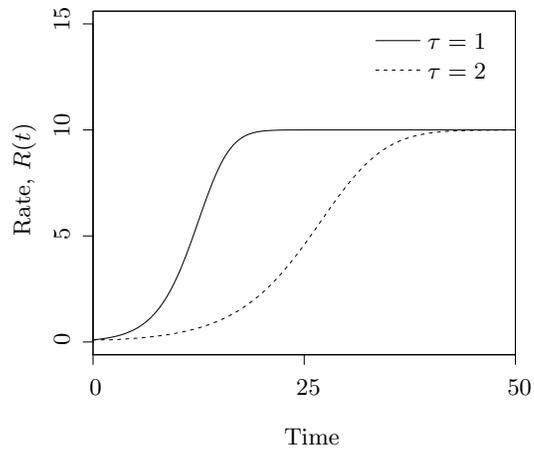}
\caption{Numerical simulations of RCP without queue feedback, highlighting the decrease in convergence rate as RTT increases. The parameter values used are $C=10$ and $a=1/e$.}
\label{fig:rocVstau}
\end{figure}

To summarize, the convergence rate $\sigma$ is given by
\begin{align}
 \sigma&=\min[\sigma_1,\sigma_2]=\sigma_2 \quad a \in (0,a^*],\\
 &=\min[\sigma_1,\sigma_3]=\sigma_3 \quad a>a^*,
\end{align}
where $a^*=1/e$. Figure \ref{fig:roc} shows the variation of convergence rate with protocol parameter $a$ for various values of $\tau$. It can observed that the rate of convergence increases with $a$ for $a<1/e$, and decreases when $a>1/e$. Also, the convergence rate is maximum at $a=(1/e)$, so the optimal value of the protocol parameter for fast convergence is $a=(1/e)$.
%%%%%%%%%%%%%%%%%%%%%%%
Also, the rate of convergence decreases with an increase in the round-trip time (see Figure \ref{fig:rocVstau}). The numerical simulations shown in Figure \ref{fig:rocVstau} are done using the computing software XPPAUT~\citep{xppaut2002}.
%%%%%%%%%%%%%%%%%%%%%%%%
%%%%%%%%%%%%%%%%%%%%%%%%%%%%%%%%%%%%%%%%%%%%%%%%%%%%%

\begin{figure}[hbtp!]
 \centering
\psfrag{R}{\hspace{-1.8cm} \small{Rate of convergence, $\sigma$}}
\psfrag{a}{{\hspace{-0.8cm}\small{Parameter, $a$}}}
\psfrag{b=0}{\hspace{-2.3mm}\small{$\beta=0$}}
\psfrag{b=0.2}{\hspace{-2.3mm}\small{$\beta=0.2$}}
\psfrag{b=0.4}{\hspace{-2.3mm}\small{$\beta=0.4$}}
\psfrag{b=0.5}{\hspace{-2.3mm}\small{$\beta=0.5$}}
\psfrag{unstable}{\small }
\psfrag{roc}{\small Hopf condition}
\psfrag{0}{\small{$0$}}
\psfrag{0.0000000}{\hspace{0.4cm}\small{$0$}}
\psfrag{0.3678794}[][]{\small{$1/e$}}
\psfrag{0.7853982}{\hspace{0.3cm}\small{$\pi/4$}}
\psfrag{1.5707963}{\hspace{0.3cm}\small{$\pi/2$}}
\psfrag{0.0}{\hspace{0.0cm}\small{$0$}}
\psfrag{0.5}{\hspace{0.0cm}\small{$0.5$}}
\psfrag{1.0}[][]{\small{$1$}}
\psfrag{1.57}{\small{$\pi/2$}}
\includegraphics[width=0.7\wdth]{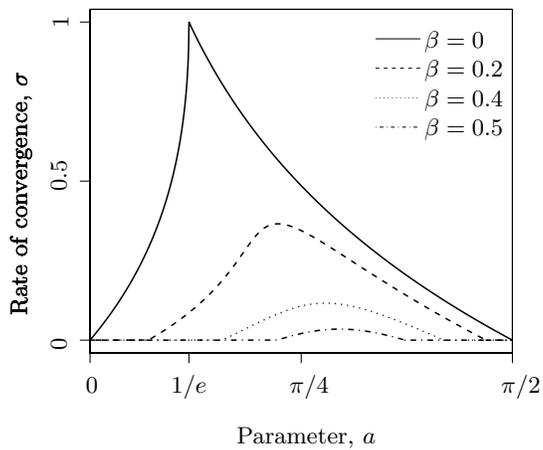}
\caption{Convergence rate to equilibrium of RCP for various values of the protocol parameter $\beta$. We can observe that the convergence rate can be improved by excluding the queue size feedback. The parameter values used are $C=10$ and $\tau=1$.}
\label{fig:roc_q}
\end{figure}

\begin{figure}[hbtp!]
\centering
\psfrag{time}{\hspace{-0.1cm}  \small Time}
\psfrag{Rate}{\hspace{-0.5cm}  \small Rate, $R(t)$}
\psfrag{withq}{\hspace{-2.5mm} \small $\beta=0.2$}
\psfrag{noq}{\hspace{-2.5mm} \small $\beta=0$}
\psfrag{0}{\small{$0$}}
\psfrag{5}{\hspace{-0.05cm}\small{$5$}}
\psfrag{10}{\hspace{-0.1cm}\small{$10$}}
\psfrag{15}{\hspace{-0.1cm}\small{$15$}}
\psfrag{20}{\hspace{-0.1cm}\small{$20$}}
\psfrag{40}{\hspace{-0.1cm}\small{$40$}}
\psfrag{15}{\hspace{-0.1cm}\small{$15$}}
\psfrag{80}{\hspace{-0.1cm}\small{$80$}}
\psfrag{0}{\small{$0$}}
\psfrag{25}{\hspace{0cm}\small{$25$}}
\psfrag{50}{\hspace{-0.1cm}\small{$50$}}
\psfrag{100}{\hspace{-0.1cm}\small{$100$}}
\psfrag{200}{\hspace{-0.1cm}\small{$200$}}
\includegraphics[width=0.7\wdth]{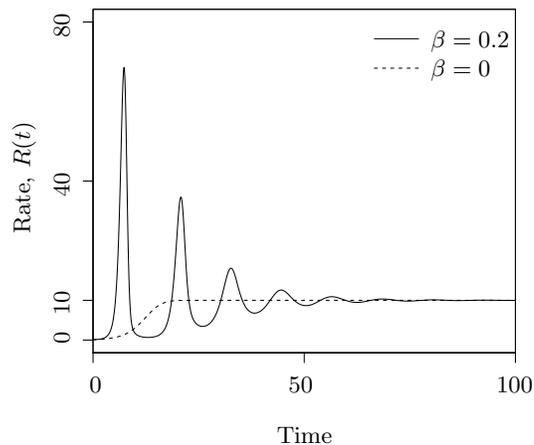}
\caption{Numerical simulations highlighting that the system can converge to its equilibrium ($R^*=10$) much faster, if the queue size feedback is not included in the RCP. The parameter values used are $C = 10$, $a=1/e$ and $\tau = 1$.} \label{fig:roc_qVsnoq_numerics}
\end{figure}

For the sake of comparison, the rate of convergence for the case $b>0$ is also shown in Figure \ref{fig:roc_q}. As can be observed from Figure \ref{fig:roc_q}, the removal of queue size feedback from RCP improves the convergence rate to equilibrium. Numerical simulations in Figure \ref{fig:roc_qVsnoq_numerics} serve to validate the analytical insights.
% Thus, to have a faster rate of convergence, the RCP with feedback based on only rate mismatch is the good design choice.

However, rather than being confined to a stable regime, it is also worth investigating the behavior of the protocol if the system becomes unstable. Therefore, in the next section, we analyze the consequences associated with the loss of stability for both the design choices.

%%%%%%%%%%%%%%%%%%%%%%%%%%%%%%%%%%%%%%%%%%%%%%HOPF BIFURCATON ANALYSIS%%%%%%%%%%%%%%%%%%%%%%%%%%%%%%%%%%%%%%%%%%%%%%%%%%%%%%%%%%%%%%%%%

\section{Local Hopf bifurcation analysis}
In the previous sections, we analyzed local stability and convergence properties of the RCP protocol. The next natural step is to investigate the dynamical behavior of the system as it transits from a stable to an unstable regime. In this section, we explore the impact of loss of local stability for both the design options, i.e., with and without queue size feedback.
% Local stability analysis retains only the linear component and ignores all
% higher order terms of the non-linear system before addressing the issue of stability. So, it looks appealing to have an analytical methodology which may allow us to capture the impact of some non-linear terms while performing a Taylor expansion of the non-linear system about its equilibrium. Local bifurcation theory is one such methodology \citep{hassard1981}. 
In the bifurcation-theoretic analysis, we have to take non-linear terms into consideration, which helps to learn additional non-linear dynamical properties of the system with and without queue size feedback.
%In this section, we investigate the bifurcation properties of RCP in the presence and absence of queue feedback. 

\subsection{With queue feedback}
\label{section:Hopf}
In the local Hopf bifurcation analysis, we use the theoretical framework of Poincar{\`e} normal forms and the center manifold theorem to investigate the type of Hopf bifurcation and the asymptotic orbital stability of the bifurcating limit cycles. For our study, we closely follow the style of analysis provided in \citep{hassard1981}.

We start by conducting a Taylor series expansion of~\eqref{eq:simpmodel_RCP} about the equilibrium,
to get
%
% Add $n$ to the equation
\begin{equation}
\begin{split}
\label{eq:e20}
\frac{d}{dt}r(t) & =  -\kappa\left( \frac{a}{\tau}r(t-\tau) + \frac{\beta}{\tau^{2}}q(t) + \frac{a}{C\tau}r(t)r(t-\tau)
 + \frac{\beta}{C\tau^{2}}r(t)q(t) \right),
\\
\frac{d}{dt}q(t) & = \kappa r(t-\tau).
\end{split}
\end{equation}
%
%
% %
% We now perform computations that will enable us to address the questions
% about the form of the bifurcating solutions of (\eqref{eq:e20}), as it transits
% from stability to instability via a Hopf bifurcation. 
%
%
At $\kappa=\kappa_c$, the system satisfies the Hopf bifurcation condition, and a further increase in $\kappa$ would drive the system into the unstable region.
Let $\kappa = \kappa_{c} + \mu$, where $\kappa_{c} = 1$, and the Hopf bifurcation takes place at
$\mu=0$. 

Consider the following autonomous delay-differential system
\begin{equation}
\label{eq:e22}
\frac{d}{dt}\boldsymbol{u}(t) = \mathcal{L_{\mu}}\boldsymbol{u}_{t}+\mathcal{F}(\boldsymbol{u}_{t},\mu)
\end{equation}
where $t>0$, $\mu$ $\in$ $\mathbb{R}$, $\tau>0$,
\[
\boldsymbol{u}_{t}(\theta) = \boldsymbol{u}(t+\theta), \qquad \boldsymbol{u}:\big[-\tau,0\big] \rightarrow \mathbb{R}^{2}, \qquad \theta \in [-\tau,0].
\]
$\mathcal{L_{\mu}}$ is a one-parameter family of continuous (bounded) linear operators. 
The operator $\mathcal{F}(\boldsymbol{u}_t,\mu)$ contains the non-linear terms. 
We assume that $\mathcal{F}$ and $\mathcal{L_{\mu}}$ depend analytically on the bifurcation parameter $\mu$ for small $|\mu|$. 
Note that \eqref{eq:e20} is of the form \eqref{eq:e22}, 
where $\boldsymbol{u} = \big[r,q\big]^{T}$, and
\begin{align*}
   \mathcal{L_{\mu}}\boldsymbol{\phi}&=\,\,\,\, \kappa\left[ \begin{array}{cc}
0 & -\frac{\beta}{\tau^2} \\
0 & 0 	\end{array} \right]\boldsymbol{\phi}(0) +
\kappa\left[ \begin{array}{cc}
-\frac{a}{\tau}  & 0 \\
1 & 0 	\end{array} \right]\boldsymbol{\phi}(-\tau),\\
\mathcal{F}(\boldsymbol{u}_t,\mu) &= -\kappa\left[ \begin{array}{cc}
\frac{a}{C\tau}r(t)r(t-\tau)
 + \frac{\beta}{C\tau^{2}}r(t)q(t) \\
0 \end{array}  \right].
\end{align*}

The objective now is to cast \eqref{eq:e22} into the form which has $\boldsymbol{u}_{t}$ instead of both $\boldsymbol{u}$ and $\boldsymbol{u}_{t}$, i.e.,
\begin{equation}
\label{eq:e23}
\frac{d}{dt}\boldsymbol{u}_{t} = \mathcal{A}(\mu)\boldsymbol{u}_{t}+\mathcal{R}\boldsymbol{u}_{t}.
\end{equation}
\hspace{-2.5pt}First,  transform the linear problem $d\boldsymbol{u}(t)/dt = \mathcal{L_{\mu}}\boldsymbol{u}_{t}$. 
By Riesz representation theorem, there exists a $2\times2$ matrix function $\eta(\cdot,\mu):[-\tau,0] \rightarrow \mathbb{R}^{2\times2}$, such that the components of $\eta$ have bounded variation and for all $\boldsymbol{\boldsymbol{\phi}}$ $\in$ $C[-\tau,0]$
\begin{align}
\mathcal{L_{\mu}}\boldsymbol{\boldsymbol{\phi}} &= \int^{0}_{-\tau}\mathrm{d}\eta(\theta,\mu)\boldsymbol{\boldsymbol{\phi}}(\theta). \notag
%\]
%
\shortintertext{In particular,}
{\color{black}\mathcal{L_{\mu}}\boldsymbol{u}_{t} } &= \int^{0}_{-\tau}\mathrm{d}\eta(\theta,\mu) \boldsymbol{u} (t+\theta). \label{eq:e24}
\intertext{Observe that}
%
%\[
\mathrm{d}\eta(\theta,\mu) &= \kappa\left[ \begin{array}{cc}
-\frac{a}{\tau}\delta(\theta +\tau) & - \frac{\beta}{\tau^{2}}\delta(\theta) \\
 \delta(\theta +\tau) & 0 	\end{array} \right]\mathrm{d}\theta
%\]
\end{align}
satisfies \eqref{eq:e24}. Here, $\delta(\theta)$ is the Dirac delta function. 
For $\boldsymbol{\boldsymbol{\phi}}$ $\in$ $C^1\big[-\tau,0\big]$, define
\begin{align}
\mathcal{A}(\mu)\boldsymbol{\phi}(\theta) &= 
	\begin{cases}
	    \begin{array}{c} \frac{\mathrm{d}\boldsymbol{\phi}(\theta)}{\mathrm{d}\theta},\\
		\int^{0}_{-\tau}\mathrm{d}\eta(s,\mu)\boldsymbol{\phi}(s) \equiv \mathcal{L_{\mu}}\boldsymbol{\phi}, \end{array}  & 
	    \begin{array}{l}  \theta \in [-\tau,0),\\ \theta = 0, \end{array}
	\end{cases} \label{eq:e25}
\intertext{and}
\mathcal{R}\boldsymbol{\phi}(\theta) &= 
	  \begin{cases}
	      \begin{array}{c} 0,\\ \mathcal{F}(\boldsymbol{\phi},\mu),\end{array} & 
	      \begin{array}{l} \theta \in [-\tau,0), \\ \theta = 0. \end{array}
	   \end{cases} \label{eq:e26}
\end{align}
As $\mathrm{d}\boldsymbol{u}_{t}/\mathrm{d}\theta=\mathrm{d}\boldsymbol{u}_{t}/\mathrm{d}t$, \eqref{eq:e22} becomes \eqref{eq:e23} as desired. Let $\boldsymbol{q}(\theta)$ be the eigenfunction for $\mathcal{A}(0)$ corresponding to $\lambda(0)$, 
\begin{equation}
 \mathcal{A}(0)\boldsymbol{q}(\theta)= i\omega_{0}\boldsymbol{q}(\theta). \label{eq:Aq}
\end{equation}
To find $\omega_{0}$ and $\boldsymbol{q}(\theta)$, let $\boldsymbol{q}(\theta)=\boldsymbol{q}_0e^{i\omega_{0}\theta}$, where $\boldsymbol{q}_0=[q_{01},q_{02}]^{T}$. Now, using \eqref{eq:Aq} and \eqref{eq:e25}, and following the style of analysis done for the equations \eqref{eq:reim} and \eqref{eq:reim1}, we get
\begin{equation}
\label{eq:e27}
\begin{aligned}
\kappa{}a &= (\omega_0\tau)\sin(\omega_0\tau), \qquad&\qquad
\kappa{}^2\beta &= (\omega_0\tau)^2\cos(\omega_0\tau), \\
\boldsymbol{q}_0 &= \left[ \begin{array}{c}1\\   -\frac{i\omega_0\tau^2}{ (\kappa{}\beta + ia\omega_0\tau )}  \end{array} \right] \equiv \left[\begin{array}{c}q_{01}\\q_{02}\end{array}\right], &
\omega_{0} &= \frac{\kappa{}}{\tau}\sqrt{\frac{ a^2 + \sqrt{ a^4 + 4\beta^2}}{2}}. %\equiv\sqrt{\frac{a^{2}_{1} + \sqrt{a^{4}_{1} + 4a^{2}_{2}}}{2}}\\
\end{aligned}
\end{equation}
Define the adjoint operator $\mathcal{A^{*}}(0)$ as
\[
\mathcal{A^{*}}(0)\boldsymbol{\alpha}(s) = \begin{cases}
\begin{array}{c}
-\frac{\mathrm{d}\boldsymbol{\alpha}(s)}{\mathrm{d}s},\\
\int^{0}_{-\tau}\mathrm{d}\eta^{T}(t,0)\boldsymbol{\alpha}(-t),\end{array} 
& \begin{array}{l}
s \in (0,\tau],\\
s = 0. \end{array}\end{cases}
\]
Note that the domains of $\mathcal{A}$ and $\mathcal{A^{*}}$ are $C^1[-\tau,0]$ and $C^1[0,\tau]$, respectively. As 
\[
\mathcal{A}(0)\boldsymbol{q}(\theta)= \lambda(0)\boldsymbol{q}(\theta)
\]
$\overline{\lambda}(0)$ is an eigenvalue for $\mathcal{A^{*}}$, and 
\[
\mathcal{A^{*}}(0)\boldsymbol{q}^*=-i\omega_{0}\boldsymbol{q}^*
\]
for some non-zero vector $\boldsymbol{q}^*$. Thus, we get
\begin{align}
 \boldsymbol{q}^* &= \Omega \left[ \begin{array}{c}1\\   \frac{\kappa{}\beta}{i\omega_0\tau^2}   \end{array}\right]   
 \equiv \left[\begin{array}{c}q^*_{01}\\q^*_{02}\end{array}\right],
\end{align}
where $\Omega$ is a non-zero scalar. For $\boldsymbol{\phi}\in{}C[-\tau,0]$ and $\boldsymbol{\psi}\in{}C[0,\tau]$, define an inner product
\begin{equation}
\label{eq:e28}
\langle\boldsymbol{\psi},\boldsymbol{\phi}\rangle=\overline{\boldsymbol{\psi}}(0)\cdot\boldsymbol{\phi}(0)
-\underset{\theta=-\tau}{\int^{0}}\underset{\zeta=0}{\int^{\theta}}\overline{\boldsymbol{\psi}}^{T}(\zeta-\theta)\mathrm{d}\eta(\theta)\boldsymbol{\phi}(\zeta)\mathrm{d}\zeta,
\end{equation}
where $\boldsymbol{a}$ $\cdot$ $\boldsymbol{b}$ means $\sum a_ib_i$. Then, $\langle\boldsymbol{\psi},\mathcal{A}\boldsymbol{\phi}\rangle = \langle\mathcal{A^{*}}\boldsymbol{\psi},\boldsymbol{\phi}\rangle$ for $\boldsymbol{\phi} \in \mathrm{Dom}(\mathcal{A})$ and $\boldsymbol{\psi}\in{}\mathrm{Dom}(\mathcal{A})$. Using \eqref{eq:e28} to find $\Omega$ such that $\left\langle \boldsymbol{q}^* , \boldsymbol{q} \right\rangle=1$, we get
\newcommand{\Omegabar}{\overline{\Omega}}
\newcommand{\qbar}{\overline{q}}
$$ 1 = \Omegabar \Big( 1 + q_{02}\qbar^*_{02}  + e^{-i\omega_0\tau}\big( \qbar^*_{02}\tau -\kappa{}a\big) \Big).$$
On simplification, we obtain
\begin{align}
 \Omega &= \bigg( 1 - i\omega_0\tau + \frac{\kappa\beta}{\kappa\beta - ia\omega_0\tau} \bigg)^{-1}. \label{eq:Omegarel}
\end{align}
Similarly, we can also verify that $\langle \boldsymbol{q}^*,\overline{\boldsymbol{q}} \rangle=0$. 
For $\boldsymbol{u}_t$, a solution of \eqref{eq:e23} at $\mu=0$, define
\[
\begin{aligned}
z(t) &= \langle{}\boldsymbol{q}^*,\boldsymbol{u}_{t}\rangle,\\
\boldsymbol{w}(t,\theta) &= \boldsymbol{u}_t(\theta)-2\mathbf{Re}\,\big( z(t)\boldsymbol{q}(\theta)\big).
\end{aligned}
\]
Then, on the manifold, $C_{0}$, $\boldsymbol{w}(t,\theta)=\boldsymbol{w}\big(z(t),\overline{z}(t),\theta\big)$, where
\begin{equation}
\label{eq:e31}
\boldsymbol{w}\big(z,\overline{z},\theta\big)=\boldsymbol{w}_{20}(\theta)\frac{z^{2}}{2}+\boldsymbol{w}_{11}(\theta)z\overline{z}+\boldsymbol{w}_{02}(\theta)\frac{\overline{z}^{2}}{2}+\cdots.
\end{equation} 
In effect, $z$ and $\overline{z}$ are local coordinates for $C_{0}$ in $C$ in the directions
of $\boldsymbol{q}^*$ and $\overline{\boldsymbol{q}}^{*}$, respectively.
%Note that $\boldsymbol{w}$ is real if $\boldsymbol{u}_t$ is real, and we deal only with real solutions.
The existence of the center manifold enables the reduction of \eqref{eq:e23} to an ordinary differential equation for a single complex variable on $C_{0}$. At $\mu=0$, we get
\begin{equation}
\begin{aligned}
\label{eq:e32}
z^{'}(t) &= \langle \boldsymbol{q}^*,\mathcal{A}\boldsymbol{u}_t+\mathcal{R}\boldsymbol{u}_t\rangle,\\
&= i\omega_{0}z(t)+\overline{\boldsymbol{q}}^{*}(0)\cdot\mathcal{F}\bigg(\boldsymbol{w}(z,\overline{z},\theta)+2\mathbf{Re}\Big( z\boldsymbol{q}(\theta)\Big)\bigg),\\
&= i\omega_{0}z(t)+\overline{\boldsymbol{q}}^{*}(0)\cdot\mathcal{F}_{0}(z,\overline{z}),
\end{aligned}
\end{equation} 
which can be abbreviated as 
\begin{equation}
\label{eq:e33}
z^{'}(t) = i\omega_{0}z(t)+g\big(z,\overline{z}\big).
\end{equation} 

The next objective is to expand $g$ in powers of $z$ and $\overline{z}$. However, we also have to determine the coefficients $\boldsymbol{w}_{ij}(\theta)$ in \eqref{eq:e31}. Once the coefficients $\boldsymbol{w}_{ij}$ are determined, the differential equation \eqref{eq:e32} for $z$ would be explicit (as abbreviated in \eqref{eq:e32}). Expanding $g(z,\overline{z})$ in powers of $z$ and $\overline{z}$, we have
\[
\begin{aligned}
g(z,\overline{z}) &= \overline{\boldsymbol{q}}^{*}(0)\cdot\mathcal{F}_{0}\big(z,\overline{z}\big),\\
&= g_{20}\frac{z^{2}}{2}+g_{11}z\overline{z}+g_{02}\frac{\overline{z}^{2}}{2}+g_{21}\frac{z^{2}\overline{z}}{2}\cdots.
\end{aligned}
\]
Following \citep{hassard1981}, we write
\[
\boldsymbol{w}^{'} = u^{'}_{t}-z^{'}\boldsymbol{q}-\overline{z}^{'}\overline{\boldsymbol{q}},
\]
and using \eqref{eq:e23} and \eqref{eq:e33} we obtain
\[
\boldsymbol{w}^{'} = \begin{cases}
\begin{array}{l}
\mathcal{A}\boldsymbol{w}-2\mathbf{Re}\left(\overline{\boldsymbol{q}}^{*}(0)\cdot\mathcal{F}_{0}\boldsymbol{q}(\theta)\right),\\
\mathcal{A}\boldsymbol{w}-2\mathbf{Re}\left(\overline{\boldsymbol{q}}^{*}(0)\cdot\mathcal{F}_{0}q(0)\right)+\mathcal{F}_{0},\end{array} & \begin{array}{l}
\theta \in [-\tau,0),\\
\theta = 0,\end{array}\end{cases}
\]
which is rewritten as
\begin{equation}
\label{eq:e34}
\boldsymbol{w^{'}} = \mathcal{A}\boldsymbol{w}+\boldsymbol{h}\big(z,\overline{z},\theta\big)
\end{equation}
using \eqref{eq:e31}, where
\begin{equation}
\label{eq:e35}
\boldsymbol{h}(z,\overline{z},\theta)=\boldsymbol{h}_{20}(\theta)\frac{z^{2}}{2}+\boldsymbol{h}_{11}(\theta)z\overline{z}+\boldsymbol{h}_{02}(\theta)\frac{\overline{z}^{2}}{2}+\cdots.
\end{equation}
On the manifold, $C_{0}$, near the origin
\[
\boldsymbol{w}^{'} = \boldsymbol{w}_{z}z^{'}+\boldsymbol{w}_{\overline{z}}\overline{z}^{'}.
\]
Using \eqref{eq:e31} and \eqref{eq:e33} to replace $\boldsymbol{w}_{z}$, $z^{'}$ (and their conjugates by their power series expansion) and equating this with \eqref{eq:e34}, we get
\begin{equation}
\label{eq:e36}
\begin{aligned}
\big(2i\omega_{0}-\mathcal{A}\big)\boldsymbol{w}_{20}(\theta) &= \boldsymbol{h}_{20}(\theta),\\
-\mathcal{A}\boldsymbol{w}_{11}(\theta) &= \boldsymbol{h}_{11}(\theta),\\
-\big(2i\omega_{0}+\mathcal{A}\big)\boldsymbol{w}_{02}(\theta) &= \boldsymbol{h}_{02}(\theta).
\end{aligned}
\end{equation}

Note that
\[
\begin{aligned}
\boldsymbol{u}_t(\theta)&=\boldsymbol{w}\big(z,\overline{z},\theta\big)+\boldsymbol{q}(\theta)z+\overline{\boldsymbol{q}}(\theta)\overline{z},\\
&=\boldsymbol{w}_{20}(\theta)\frac{z^{2}}{2}+\boldsymbol{w}_{11}(\theta)z\overline{z}+\boldsymbol{w}_{02}(\theta)\frac{\overline{z}^{2}}{2} 
+\boldsymbol{q}_0e^{i\omega_{0}\theta}z+\overline{\boldsymbol{q}}_{0}e^{-i\omega_{0}\theta}\overline{z}+\cdots,
\end{aligned}
\]
from which we obtain $\boldsymbol{u}_t(0)$ and $\boldsymbol{u}_t(-\tau)$.
%We have actually looked ahead and 
We need to find the coefficients of $z^{2}$, $z\overline{z}$, $\overline{z}^{2}$, $z^{2}\overline{z}$. Hence, we only keep these relevant terms in the expansions as follows:
%Note that we have only two non-linear terms, $r_t(0)q_t(0)$ and $r_t(0)r_t(-\tau)$
\begin{equation}\small
\begin{aligned}
 r_t(0)q_t(0) &= q_{02}z^{2}+\overline{q}_{02}\overline{z}^{2}+\big(\overline{q}_{02}+q_{02}\big)z\overline{z} +\Bigg( \frac{w_{201}(0)\overline{q}_{02}}{2}+w_{111}(0)q_{02}+w_{112}(0)\\
 &\quad +\frac{w_{202}(0)}{2} \Bigg)z^{2}\overline{z}+\cdots,\\
r_t(0)r_t(-\tau) &= e^{-i\omega_{0}\tau}z^{2}+e^{i\omega_{0}\tau}\overline{z}^{2}+\left(e^{i\omega_{0}\tau}+e^{-i\omega_{0}\tau}\right)z\overline{z} 
+\Bigg( \frac{w_{201}(0)e^{i\omega_{0}\tau}}{2}
\\ &\quad +w_{111}(0)e^{-i\omega_{0}\tau} + w_{111}(-\tau)+\frac{w_{201}(-\tau)}{2} \Bigg)\ z^{2}\overline{z}+\cdots,
\end{aligned}
\label{eq:gnonlinearterms}
\end{equation}
where $[w_{ij1},w_{ij2}]^{T}=\boldsymbol{w}_{ij}$.
Recall that,
\begin{equation}
g(z,\overline{z})=\overline{\boldsymbol{q}}^{*}(0)\cdot\mathcal{F}_{0}(z,\overline{z}),
\label{eq:gexpr1}\end{equation}
where $[\mathcal{F}_{01},\mathcal{F}_{02}]^{T}=\mathcal{F}_{0}\,$, and 
\begin{equation}
g(z,\overline{z}) = g_{20}\frac{z^{2}}{2}+g_{11}z\overline{z}+g_{02}\frac{\overline{z}^{2}}{2}+g_{21}\frac{z^{2}\overline{z}}{2}\cdots.
\label{eq:gexpr2}\end{equation}
\newcommand{\zbar}{\overline{z}}
Comparing \eqref{eq:gexpr1} and \eqref{eq:gexpr2}, and using \eqref{eq:gnonlinearterms}, we get
\begin{align*}
g_{20} &=  -2\overline{\Omega}\kappa{} \left( \frac{a}{C\tau}e^{-i\omega_0\tau} + \frac{\beta}{C\tau^2}q_{02} \right), \notag  \notag \\
g_{11} &= -\overline{\Omega}\kappa{}\,\bigg( \frac{a}{C\tau}\left( e^{i\omega_0\tau} + e^{-i\omega_0\tau} \right) + \frac{\beta}{C\tau^2}\left(\qbar_{02} + q_{02} \right) \bigg), \notag \\
g_{02} &= -2\overline{\Omega}\kappa{}\,\biggl( \frac{a}{C\tau}e^{i\omega_0\tau}  + \frac{\beta}{C\tau^2}\qbar_{02} \biggr), \notag 
\end{align*}
\begin{align*}
g_{21} &= \overline{\Omega}\kappa{}\,\Biggl( -\frac{a}{C\tau}\Bigl( w_{201}(-\tau) +   2w_{111}(-\tau) +  w_{201}(0)e^{i\omega_0\tau}
		    + 2w_{111}(0)e^{-i\omega_0\tau} \Bigr) \notag \\
		    &\quad-\frac{\beta}{C\tau^2}\Bigl( \qbar_{02}w_{201}(0) +   2q_{02}w_{111}(0)  + w_{202}(0) + 2w_{112}(0)   \Bigr) \Biggr).\notag 
\end{align*}
On further simplification, we obtain
\begin{align*}
 g_{20}    &=  \frac{2i\Omegabar\omega_0}{C},   \qquad\qquad
 \qquad\qquad g_{11}   = 0, \qquad\qquad  
  \qquad\qquad g_{02}   = \frac{-2i\Omegabar\omega_0}{C}, \label{eq:gsimplified} \\%\label{eq:e39} 
  g_{21}   &= -\frac{\Omegabar}{C\tau}\Biggl( i\omega_0\tau{}w_{201}(0) + \kappa{}aw_{201}(-\tau)    +\frac{\kappa{}\beta{}w_{202}(0)}{\tau}  
		     -2i\omega_0\tau{}w_{111}(0) \\ \nonumber
		       & \quad + 2\kappa{}aw_{111}(-\tau) +\frac{2\kappa{}\beta{}w_{112}(0)}{\tau}   \Biggr). 
\end{align*}
For the expression of $g_{21}$, we still need to evaluate $\boldsymbol{w}_{11}(0)$, $\boldsymbol{w}_{11}(-\tau)$, $\boldsymbol{w}_{20}(0)$ and $\boldsymbol{w}_{20}(-\tau)$. Now for $\theta$ $\in$ $[-\tau,0)$ 
\[
\begin{aligned}
\boldsymbol{h}(z,\overline{z},\theta) &=  -2\mathbf{Re}\,\Big(\overline{\boldsymbol{q}}^{*}(0)\cdot\mathcal{F}_{0}\boldsymbol{q}(\theta)\Big),\\
&= -2\mathbf{Re}\,\Big(g(z,\overline{z})\boldsymbol{q}(\theta)\Big),\\
%&= -g\left(z,\overline{z}\right)\boldsymbol{q}(\theta)-\overline{g}(z,\overline{z})\overline{\boldsymbol{q}}(\theta)\\
&= \left(g_{20}\frac{z^{2}}{2}+g_{11}z\overline{z}+g_{02}\frac{\overline{z}^{2}}{2}\right)\boldsymbol{q}(\theta)
-\left(\overline{g}_{20}\frac{z^{2}}{2}+\overline{g}_{11}z\overline{z}+\overline{g}_{02}\frac{\overline{z}^{2}}{2}\right)\overline{\boldsymbol{q}}(\theta),
\end{aligned}
\]
which, when compared with \eqref{eq:e35}, gives
\[
\begin{aligned}
\boldsymbol{h}_{20}(\theta)&=-g_{20}\boldsymbol{q}(\theta)-\overline{g}_{02}\overline{\boldsymbol{q}}(\theta),\\
\boldsymbol{h}_{11}(\theta)&=-g_{11}\boldsymbol{q}(\theta)-\overline{g}_{11}\overline{\boldsymbol{q}}(\theta) = 0. \\
\end{aligned}
\]
From \eqref{eq:e25} and \eqref{eq:e36}, we get
\begin{align}
\boldsymbol{w}^{'}_{20}(\theta) &= 2i\omega_{0}\boldsymbol{w}_{20}(\theta) + 
g_{20}\boldsymbol{q}(\theta)+\overline{g}_{02}\overline{\boldsymbol{q}}(\theta), \label{eq:e41} \\
\boldsymbol{w}^{'}_{11}(\theta) &= 0. \label{eq:e42}
\end{align}
Solving the differential equations \eqref{eq:e41} and \eqref{eq:e42}, we obtain
\begin{align}
\boldsymbol{w}_{20}(\theta) &= -\frac{g_{20}}{i\omega_{0}}\boldsymbol{q}_0e^{i\omega_{0}\theta}-\frac{\overline{g}_{02}}{3i\omega_{0}}\overline{\boldsymbol{q}}_{0}e^{-i\omega_{0}\theta}+\boldsymbol{e}e^{2i\omega_{0}\theta}, \label{eq:e43} \\
\boldsymbol{w}_{11}(\theta) &= \boldsymbol{f}, \label{eq:e44}
\end{align}
for some $\boldsymbol{e}=[e_{1},e_{2}]^{T}$ and $\boldsymbol{f}=[f_{1},f_{2}]^{T}$.\\ 
From $\boldsymbol{h}(z,\overline{z},0) =  -2\mathbf{Re}\Big(\overline{\boldsymbol{q}}^{*}(0)\cdot\mathcal{F}_{0}\boldsymbol{q}(0)\Big)+\mathcal{F}_{0}$, we obtain
\[
\begin{aligned}
\boldsymbol{h}_{20}(0)=& -g_{20}\boldsymbol{q}(0)-\bar{g}_{02}\overline{\boldsymbol{q}}(0)
 - \left[\begin{array}{c}2\kappa\left(\frac{a}{C\tau}e^{-i\omega_{0}\tau}+\frac{\beta}{c\tau^2}q_{02}\right)\\0\end{array}\right],\\
\boldsymbol{h}_{11}(0)=& -g_{11}q(0)-\bar{g}_{11}\overline{\boldsymbol{q}}(0)
 - \left[\begin{array}{c}\frac{a\kappa}{C\tau}\big(e^{i\omega_{0}\tau}+e^{-i\omega_{0}\tau}\big)+\frac{\beta\kappa}{C\tau^2}\left(\overline{q}_{02}+q_{02}\right)\\0\end{array}\right].
\end{aligned}
\]
On further simplifying the above equations, we get
\begin{equation}
 \boldsymbol{h}_{20}(0) = -\frac{2i\omega_0}{C} \left( \Omegabar\boldsymbol{q}_0 + \Omega\boldsymbol{\qbar}_0 -  \left[ \begin{array}{c} 1 \\ 0  \end{array} \right]\, \right),\label{eq:h200}
\end{equation}
\begin{equation}
 \hspace{-45mm}\boldsymbol{h}_{11}(0) = 0.\label{eq:h110}
\end{equation}
From \eqref{eq:e25}, \eqref{eq:e36}, \eqref{eq:h200} and \eqref{eq:h110}, we get
\newcommand{\Dbar}{\overline{D}}
\begin{align}
 \left[\!\! \begin{array}{c} 2i\omega_0 w_{201}(0) +\frac{\kappa{}a}{\tau}w_{201}(-\tau) +\frac{\kappa{}\beta}{\tau^2}w_{202}(0) \\ 2i\omega_0 w_{202}(0) - \kappa{} w_{201}(-\tau)  \end{array}  \!\!\right] 
	&= -\frac{2i\omega_0}{C} \left( \Omegabar\boldsymbol{q}_0 + \Omega\boldsymbol{\qbar}_0 -  \left[ \begin{array}{c} 1 \\ 0  \end{array} \right] \right), \label{eq:e45} \\
   \left[ \begin{array}{c}
		-\frac{\kappa{}a}{\tau}w_{111}(-\tau) -\frac{\kappa{}\beta}{\tau^2}w_{112}(0)   \\
		\kappa{}w_{111}(-\tau)  	
    \end{array} \right]  &= 0. \label{eq:e46}
\end{align}
Substituting the expression for $w_{ijk}(x), x\in[-\tau,0]$ (from \eqref{eq:e43} and \eqref{eq:e44}) in the equations \eqref{eq:e45} and \eqref{eq:e46}, 
and solving for $e_{1}$, $e_{2}$, $f_{1}$ and $f_{2}$, we get
\begin{align}
\boldsymbol{e} \equiv \left[\begin{array}{c}e_{1}\\e_{2}\end{array}\right] &= \frac{2\tau^2}{ CA_1 }
  \left[  \begin{array}{c}
     2    \\
       i\omega_0q_{02}^2   \end{array} \right], &
\boldsymbol{f} \equiv \left[\begin{array}{c}f_{1}\\f_{2}\end{array}\right] &= 0 \label{eq:e48},
\end{align}
where $$A_1 =  4\tau^2  + \kappa{}q_{02}^2 \left( \beta + 2ia\omega_0\tau  \right).  $$

Using \eqref{eq:e43}, \eqref{eq:e44} and \eqref{eq:e48}, we can find the solutions for $\boldsymbol{w}_{11}(0)$, $\boldsymbol{w}_{11}(-\tau)$, $\boldsymbol{w}_{20}(0)$ and $\boldsymbol{w}_{20}(-\tau)$. Using these terms, we evaluate the 
% We now substitute the expressions for $\boldsymbol{e}$ and $\boldsymbol{f}$, in (\eqref{eq:e43}) and (\eqref{eq:e44}), 
% to obtain the expressions for $\boldsymbol{w}_{11}(0)$, $\boldsymbol{w}_{11}(-\tau)$, $\boldsymbol{w}_{20}(0)$ and $\boldsymbol{w}_{20}(-\tau)$. 
% Using these we finally evaluate $g_{21}$. Simplifying the 
expression for $g_{21}$ as
\begin{align}
  g_{21} &=     \frac{2i\Omegabar\omega_0}{C^2}\bigg( \frac{2\Omega}{3}  -\frac{A_2}{A_1} \bigg),  \label{eq:g21simple}
\end{align}
where $$A_2= 2\tau^2  + \kappa{}q_{02}^2 \left( \beta + 2ia\omega_0\tau  \right).$$ 

Finally, we derived expressions for all the quantities required to compute the type of Hopf
bifurcation and the stability of the bifurcating limit cycles. Now, we can proceed with the analysis by finding out the values of $\mu_{2}$ and $\beta_{2}$ \citep{hassard1981}
\begin{equation}
\label{eq:e50}
\begin{aligned}
\mu_{2} &= \frac{ -\mathbf{Re}\Big(c_{1}(0)\Big) }{ \alpha^{'}(0) }, \qquad& \qquad
\beta_{2} &= 2\mathbf{Re}\Big(c_{1}(0)\Big),
\end{aligned}
\end{equation}
where
\begin{equation}
\label{eq:e51}
c_{1}(0) = \frac{i}{2\omega_{0}}\left(g_{20}g_{11}-2|g_{11}|^{2}-\frac{1}{3}|g_{02}|^{2}\right)+\frac{g_{21}}{2},
\end{equation}
\\
\begin{equation}
\hspace{-3.2cm} \alpha^{'}(0) = \mathbf{Re}\,\big(d\lambda/d\kappa\big)_{\kappa=\kappa_c}.
\end{equation}
\\
We now state the
definitions that will enable us to investigate the nature of the Hopf bifurcation.
\begin{itemize}[noitemsep]
 \item The Hopf bifurcation is \textit{super-critical} if $\mu_2>0$, and is \textit{sub-critical} if $\mu_2<0$.
 \item The periodic oscillations are \textit{asymptotically orbitally stable} if $\beta_{2}<0$ and unstable if  $\beta_{2}>0$.
\end{itemize}
%
%
%We now derive the expression for $\mathbf{Re}\big(c_1(0)\big)$.
Substituting the expressions for the coefficients $g_{20}$, $g_{11}$, $g_{02}$ and $g_{21}$ in \eqref{eq:e51}, we obtain
\begin{align}
 c_{1}(0)&= \frac{-i\Omegabar\omega_0A_2}{C^2A_1}, \notag \\
	  &=\frac{-i\omega_0(\beta + ia\omega_0\tau)}{C^2\big((1 + i\omega_0\tau)(\,\beta + ia\omega_0\tau) + \beta\big) }\cdot
	      \frac{   2\kappa{}(\,\beta + ia\omega_0\tau )^2 -  \omega_0^2\tau^2(\,\beta +2ia\omega_0\tau)  }
  {4\kappa{}(\,\beta + ia\omega_0\tau )^2  - \omega_0^2\tau^2(\,\beta +2ia\omega_0\tau) }.
\end{align}
We substitute the expressions for $\Omega$, $a$ and $\beta$ in terms of $\omega_0\tau$ from \eqref{eq:e27} and \eqref{eq:Omegarel} to get
\begin{align*}
    c_{1}(0)&= -\frac{i\omega_0}{C^2}\cdot\frac{   4e^{3i\omega_0\tau} -  3e^{2i\omega_0\tau} + 1  }
  { \big( (3 + 2i\omega_0\tau)e^{i\omega_0\tau} + e^{-i\omega_0\tau} \big) \big( 8e^{2i\omega_0\tau} -  3e^{i\omega_0\tau} + e^{-i\omega_0\tau} \big) }.
\end{align*}
\\
\newcommand{\myp}{\Theta}
Let  $\myp=\omega_0\tau$, and taking the real part of the above expression, we have %the real part of the above term as
{\small \begin{align*}
 \mathbf{Re}\big(c_1(0)\big) &=\frac{2\myp}{C^2\tau \left( \Big(8\cos(3\myp) - 3\cos(2\myp) + 1\Big)^2 + \Big(8\sin(3\myp) -3\sin(2\myp)\Big)^2 \right)} \\
 & \quad \times \frac{1}{\big(3+\cos(2\myp)\big)^2 + \big(2\myp-\sin(2\myp)\big)^2}\times\Bigg( 4\sin(5\myp) - 3\sin(4\myp) \\
 & \quad - 24\sin(3\myp) + 42\sin(2\myp) + 4\sin(\myp)   - 12\myp\,\big( 2\cos(3\myp)  - \cos(2\myp)\\
 & \quad - 6\cos(\myp) +  7 \big) \Bigg)
  . 
\end{align*}}

\begin{figure}[hbtp!]   
\newcommand{\mypsfrag}[2]{\psfrag{#1}{\hspace{-4pt}\scriptsize$#1$}\psfrag{#2}{\hspace{-4pt}\scriptsize$#2$}}
\centering
\mypsfrag{-0.6}{-0.2}\mypsfrag{0.0}{0.2}
\psfrag{0}{\scriptsize$0$}
\psfrag{0.0}{\scriptsize$0$}
\psfrag{-0.4}{\scriptsize$ $}
\psfrag{-0.8}{\scriptsize$ $}
\psfrag{pi/4}{\scriptsize $\pi/4$}
\psfrag{pi/2}{\scriptsize $\pi/2$}
\psfrag{th}{\scriptsize $\myp_h$}
\psfrag{theta}[t][b]{\hspace{-0.0mm}\footnotesize $\myp$}
%\psfrag{M}[b][t]{\hspace{0mm}\footnotesize $\mu_2$}
\psfrag{M}{\hspace{0mm}\footnotesize $\mu_{2}$}
\psfrag{B}{\hspace{0mm}\footnotesize $\beta_{2}$}
\includegraphics[height=4.5in,width=1.7in,angle=-90]{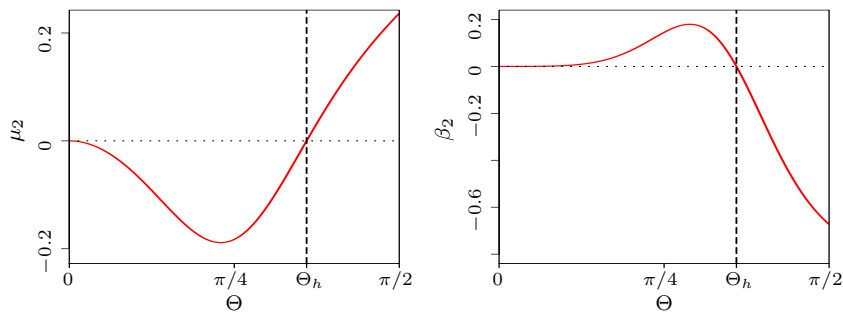}
\caption{ Variation in $\mu_2$ and $\beta_{2}$ as $\myp=\kappa\sqrt{\Big(a^{2} + \sqrt{a^{4} + 4\beta^{2}}\Big)/2}$ is varied. Observe that $\mu_2$ and
$\beta_{2}$ changes the sign at $\myp=\myp_h = 1.1297$.
Thus, the system undergoes a sub-critical Hopf bifurcation ($\mu_2<0$) and the emerging limit cycles are unstable $(\beta_{2}>0)$ for $\myp<\myp_h$. 
Whereas, for $\myp>\myp_h$, the Hopf bifurcation is super-critical and the limit cycles are asymptotically, orbitally stable.}
\label{fig:bifurc_mu2_theta}
\end{figure}

Next, we compute the value of $\alpha^{'}(0)$. 
Differentiating the characteristic equation~\eqref{eq:charwithBeta} with respect to $\kappa$, we get\\
{\small\begin{align*}
 \alpha'(0) &= \mathbf{Re}\left(\frac{d\lambda}{d\kappa}\right)\Bigg|_{\kappa=\kappa_c}\\
 &= \frac{\myp}{\kappa\tau}\cdot\frac{  4\myp\Big( 1 + \sin^2(\myp)\Big) }{ \Big(3+\cos(2\myp)\Big)^2 + \Big(2\myp - \sin(2\myp)\Big)^2 } > 0 .
% \shortintertext{and} 
%  \omega'(0) = \mathrm{Im}\left(\frac{d\lambda}{d\kappa}\right)\Bigg|_{\kappa=\kappa_c}  &= \frac{\myp}{\kappa\tau}\cdot\frac{   8 + 2\myp\sin(2\myp) + 6\cos(2\myp)  }
%  { \Big(3+\cos(2\myp)\Big)^2 + \Big(2\myp - \sin(2\myp)\Big)^2 }.
\end{align*}}
\hspace{-3pt}Therefore, if $\mathbf{Re}\,\big(c_{1}(0)\big)>0$, then $\mu_{2}<0$ (sub-critical), and $\beta_{2}>0$ (unstable limit cycles). Similarly, if  $\mathbf{Re}\,\big(c_{1}(0)\big)<0$, then $\mu_{2}>0$ and $\beta_{2}<0$, which implies that the Hopf bifurcation is super-critical, and the emerging limit cycles are orbitally asymptotically stable. In Figure \ref{fig:bifurc_mu2_theta}, we plot the variation of $\mu_2$ and $\beta_{2}$ as $\myp$ is varied.
\\
We draw the following inferences from the results of Hopf bifurcation analysis.\\

\noindent\textit{Remarks:}
\begin{itemize}[noitemsep]
 \item The nature of Hopf bifurcation and the stability of limit cycles explicitly depend on $\myp$, where
 $$\myp = \kappa\sqrt{\dfrac{a^{2} + \sqrt{a^{4} + 4\beta^{2}}}{2}}.$$
  \item Link capacity ($C$) and round-trip time ($\tau$)  do not affect the sign of $\mathbf{Re}\,\big(c_{1}(0)\big)$. Hence, the type of Hopf bifurcation is independent of these system parameters.
  \item From Figure \ref{fig:bifurc_mu2_theta}, we can note that for small values of $\myp$, we have $\mu_2<0$ and $\beta_{2}>0$, and hence the bifurcation is sub-critical and the limit cycles are unstable. 
Also, note that the criticality of the bifurcation changes from sub-critical to super-critical at $\myp_h = 1.1297$. Therefore, for $\myp>\myp_h$, the bifurcating limit cycles are asymptotically orbitally stable.
\end{itemize}
%
% The expressions for period $\mathcal{P}\left(\epsilon\right)$ and Floquet exponent $\mathcal{B}\left(\epsilon\right)$ are 
% \begin{equation}
% \label{eq:e52}
% \begin{aligned}
% \mathcal{P}\left(\epsilon\right)&=\frac{2\pi}{\omega_{0}} \bigg( 1+\epsilon^{2}\mathcal{T}_{2}+\mathcal{O}\left(\epsilon^{4}\right) \bigg) \qquad&\qquad
% \mathcal{T}_{2}&=-\frac{\mathrm{Im}\Big(c_{1}(0)\Big)+\mu_{2}\omega^{'}(0)}{\omega_{0}}\\
% \mathcal{B}\left(\epsilon\right) &=\beta_{2}\epsilon^{2}+\mathcal{O}\left(\epsilon^{4}\right) &
% \epsilon &= \sqrt{\frac{\mu}{\mu_{2}}}.
% \end{aligned}
% \end{equation}
%\paragraph{Numerical computations}

\textit{ }\\

To validate the analysis, we now perform some numerical computations for the RCP model with queue feedback. The network parameters $C$ and $\tau$ are set to $1$. We consider the following two cases.\\

\begin{figure}[hbtp!]   
\newcommand{\mypsfrag}[1]{\psfrag{#1}{\scriptsize$#1$}}
\centering
\mypsfrag{0.85}\mypsfrag{0.90}\mypsfrag{0.95}\mypsfrag{1.00}\mypsfrag{1.05}\mypsfrag{1.10}
\mypsfrag{0}\mypsfrag{1}\mypsfrag{4}\mypsfrag{8}
\psfrag{R}{\hspace{-3.5mm}\footnotesize Rate,\, $R(t)$}
\psfrag{kappa}[c]{\hspace{6mm}\footnotesize {Bifurcation parameter,} $\kappa$}
\psfrag{a0.7}[t]{\hspace{10mm}\footnotesize (a) $a=0.75,\,\,\beta=0.518$}
\psfrag{a1.3}[t]{\footnotesize (b) $a=1.25,\,\,\beta=0.454$}
\psfrag{1.04}{\scriptsize$1.05$}
\includegraphics[height=4.5in,width=1.7in,angle=-90]{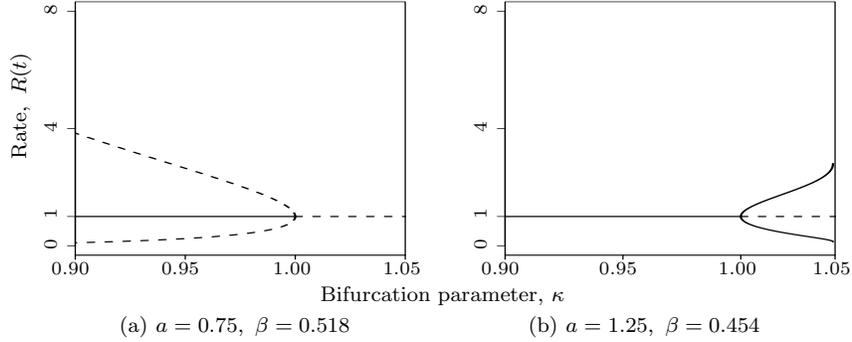}
\caption{ \textit{Bifurcation diagram} for RCP model with queue size feedback. The solid line denotes an attractor and the dashed line a repeller.
Observe that in case (a), the system undergoes a local \textit{sub-critical} Hopf, 
while in case (b), the system undergoes a local \textit{super-critical} Hopf bifurcation.
The parameters $a$ and $\beta$ are chosen so that the system is at the Hopf condition for $\kappa=1$. The other parameter values are $C=1$ and $\tau=1$. }
\label{fig:bifurc_super_sub}
\end{figure}

\noindent\textit{(i) $a=0.75$ (sub-critical)}: Let us consider $\beta=0.518$ that satisfies the Hopf condition at $\kappa=1$. For these parameter values, we obtain $\mu_2 = -0.1263$ and $\mathcal{B}_2=0.1775$. Therefore, as per the above analytical characterization, the bifurcation is sub-critical Hopf and leads to unstable limit cycles. We use DDE-Biftool~\citep{ddetool1,ddetool2} to plot the bifurcation diagram, see  Figure \ref{fig:bifurc_super_sub}(a).
As expected from the analysis, Figure \ref{fig:bifurc_super_sub}(a) shows that the system undergoes a sub-critical Hopf bifurcation. To illustrate the occurrence of a sub-critical Hopf, we present some numerical simulations (using XPPAUT) in Figure \ref{fig:sub_rcp_withq}. Considering the initial condition $R_0=1.03$ and $\kappa=0.95$, the system converges to the stable equilibrium $R^*=1$ (see Figure \ref{fig:sub_rcp_withq}(a)). Whereas, after the bifurcation, i.e., for $\kappa > \kappa_c$, the previously stable fixed point at $1$ becomes unstable and also the solution would eventually jump to infinity as shown in Figure \ref{fig:sub_rcp_withq}(b).  \\

\newcommand{\subwdth}{0.5\textwidth}

\begin{figure}[hbtp!]
\centering
\psfrag{time}{\hspace{-0.1cm}  \scriptsize Time}
\psfrag{Rate}{\hspace{-0.5cm}\vspace{2cm}\scriptsize Rate, $R(t)$}
\psfrag{0}{\hspace{-0.04cm}\scriptsize{$0$}}
\psfrag{0.0}{\hspace{-0.04cm}\scriptsize{$0.0$}}
\psfrag{5}{\hspace{-0.05cm}\scriptsize{$5$}}
\psfrag{10}{\hspace{-0.1cm}\scriptsize{$10$}}
\psfrag{15}{\hspace{-0.1cm}\scriptsize{$15$}}
\psfrag{20}{\hspace{-0.1cm}\scriptsize{$20$}}
\psfrag{25}{\hspace{0cm}\scriptsize{$25$}}
\psfrag{50}{\hspace{-0.1cm}\scriptsize{$50$}}
\psfrag{100}{\hspace{-0.1cm}\scriptsize{$100$}}
\psfrag{200}{\hspace{-0.1cm}\scriptsize{$200$}}
\psfrag{0.5}{\hspace{-0.1cm}\scriptsize{$0.5$}}
\psfrag{1.0}{\hspace{-0.1cm}\scriptsize{$1.0$}}
\psfrag{1.5}{\hspace{-0.1cm}\scriptsize{$1.5$}}
\psfrag{150}{\hspace{-0.1cm}\scriptsize{$150$}}
\psfrag{300}{\hspace{-0.1cm}\scriptsize{$300$}}
\psfrag{400}{\hspace{-0.1cm}\scriptsize{$400$}}
\psfrag{-800}{\hspace{-0.1cm}\scriptsize{$-800$}}
\psfrag{-400}{\hspace{-0.1cm}\scriptsize{$-400$}}
\psfrag{-300}{\hspace{-0.1cm}\scriptsize{$-300$}}
\psfrag{-600}{\hspace{-0.1cm}\scriptsize{$-600$}}
\psfrag{125}{\hspace{-0.1cm}\scriptsize{$125$}}
\psfrag{250}{\hspace{-0.1cm}\scriptsize{$250$}}
\psfrag{60}{\hspace{-0.1cm}\scriptsize{$60$}}
\psfrag{120}{\hspace{-0.1cm}\scriptsize{$120$}}
\begin{tabular}{c}
\subfloat[ $\kappa = 0.95$]{\includegraphics[width=\subwdth]{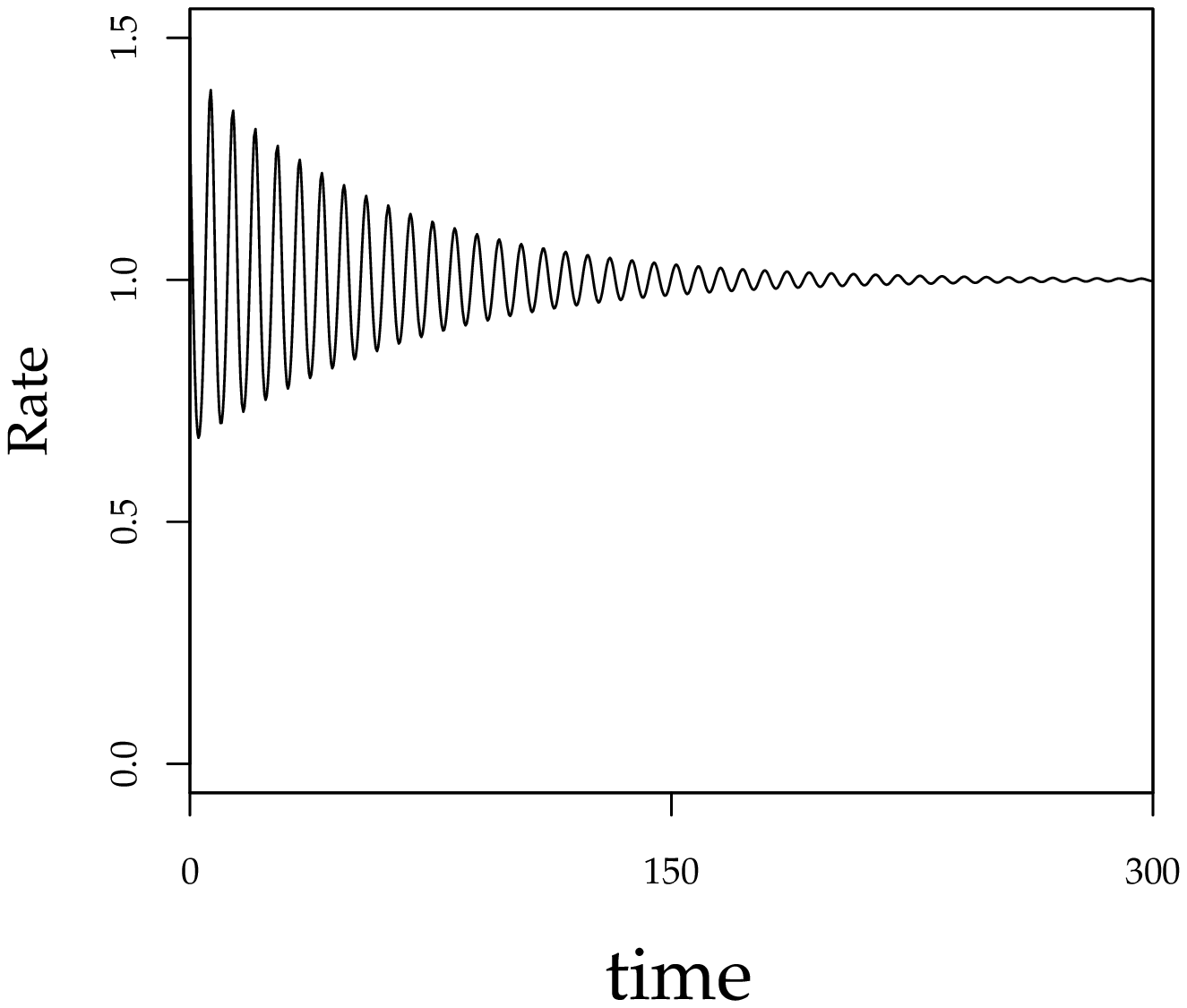}}\hspace{-5mm}
% \subfloat[ $\kappa = 0.97, R_0=2$]{\includegraphics[width=\subwdth]{tompecssub2.eps}}\hspace{-5mm}
\subfloat[ $\kappa = 1.05$]{\includegraphics[width=\subwdth]{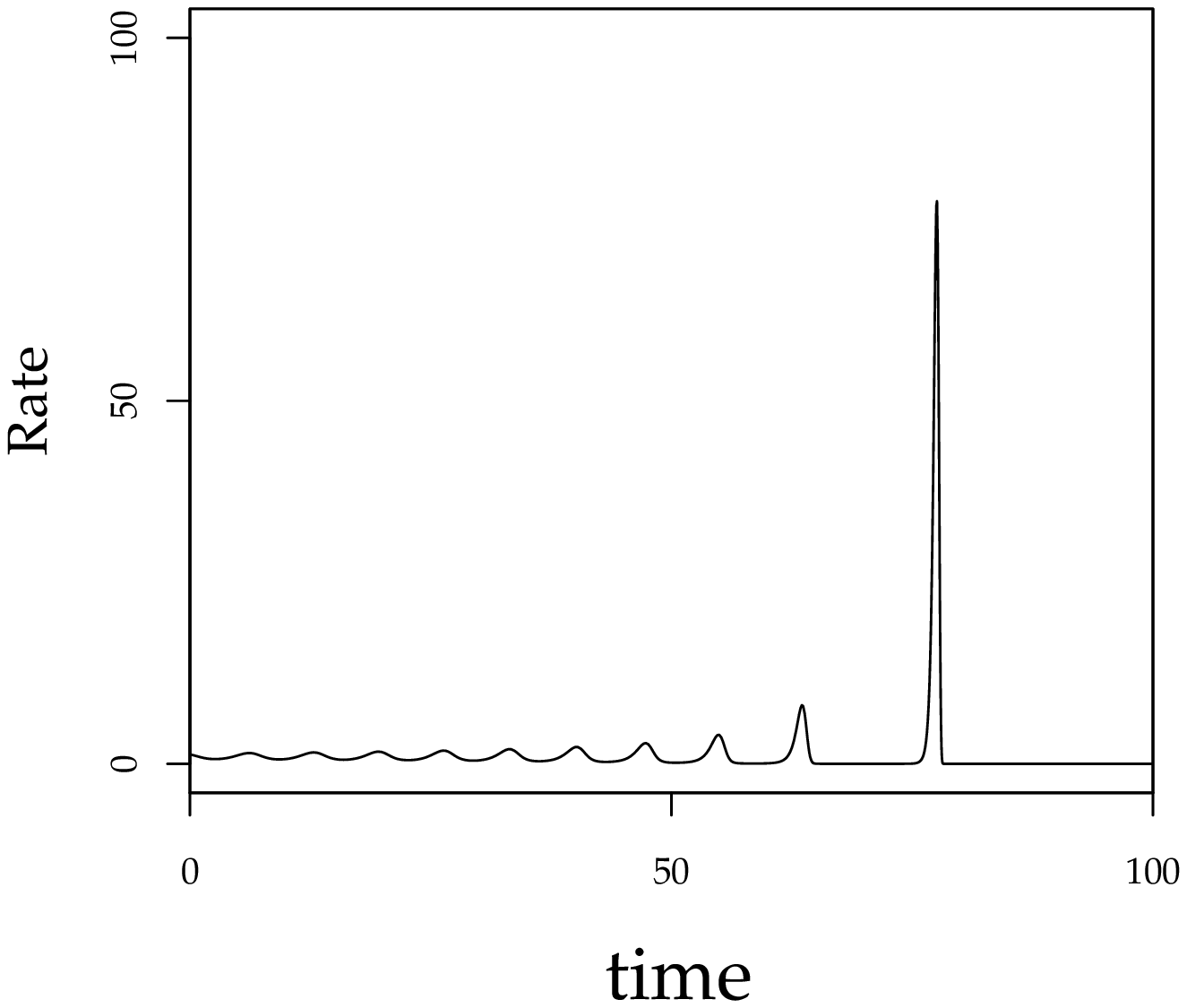}}
\end{tabular}
\caption{Numerical simulations highlighting that the system undergoes a sub-critical Hopf for the parameter values $a=0.75$, $\beta=0.518$, $C = 1$ and $\tau = 1$. } \label{fig:sub_rcp_withq}
\end{figure}

%as shown in Fig. Figure \ref{fig:sub_rcp_withq}(b), the system blows up for $R_0=2$ due to the presence of an unstable limit cycle in the neighborhood and a distant attractor at infinity. 

\noindent\textit{(ii) $a=1.25$ (super-critical)}: In this case, for the system to be at the Hopf condition with $\kappa=1$, we take $\beta=0.454$. Using these values, we calculate $\mu_2 =0.1054$ and $\mathcal{B}_2=-0.3068$. implying that the local Hopf bifurcation is super-critical and the emerging limit cycles are asymptotically orbitally stable. The bifurcation diagram in Figure \ref{fig:bifurc_super_sub}(b) shows the emergence of stable limit cycles, which corroborates the analysis. The numerical simulation shown in  Figure \ref{fig:sup_rcp_withq}(a) illustrates that the system is locally stable for $\kappa<1$. But, after the occurrence of bifurcation, the system converges to a stable limit cycle (Figure \ref{fig:sup_rcp_withq}(b)).\\
\newcommand{\supwdth}{0.5\textwidth}
\begin{figure}[hbtp!]
\centering
\psfrag{time}{\hspace{-0.1cm}  \scriptsize Time}
\psfrag{Rate}{\hspace{-0.5cm}\vspace{2cm}\scriptsize Rate, $R(t)$}
\psfrag{0}{\footnotesize{$0$}}
\psfrag{5}{\hspace{-0.05cm}\scriptsize{$5$}}
\psfrag{10}{\hspace{-0.1cm}\scriptsize{$10$}}
\psfrag{15}{\hspace{-0.1cm}\scriptsize{$15$}}
\psfrag{20}{\hspace{-0.1cm}\scriptsize{$20$}}
\psfrag{25}{\hspace{0cm}\scriptsize{$25$}}
\psfrag{50}{\hspace{-0.1cm}\scriptsize{$50$}}
\psfrag{100}{\hspace{-0.1cm}\scriptsize{$100$}}
\psfrag{200}{\hspace{-0.1cm}\scriptsize{$200$}}
\psfrag{0.0}{\hspace{-0.1cm}\scriptsize{$0.0$}}
\psfrag{0.5}{\hspace{-0.1cm}\scriptsize{$0.5$}}
\psfrag{1.0}{\hspace{-0.1cm}\scriptsize{$1.0$}}
\psfrag{1.5}{\hspace{-0.1cm}\scriptsize{$1.5$}}
\psfrag{3}{\hspace{-0.1cm}\scriptsize{$3$}}
\psfrag{1}{\hspace{-0.1cm}\scriptsize{$1$}}
\psfrag{2}{\hspace{-0.1cm}\scriptsize{$2$}}
\begin{tabular}{c}
\subfloat[ $\kappa = 0.95$]{\includegraphics[width=\supwdth]{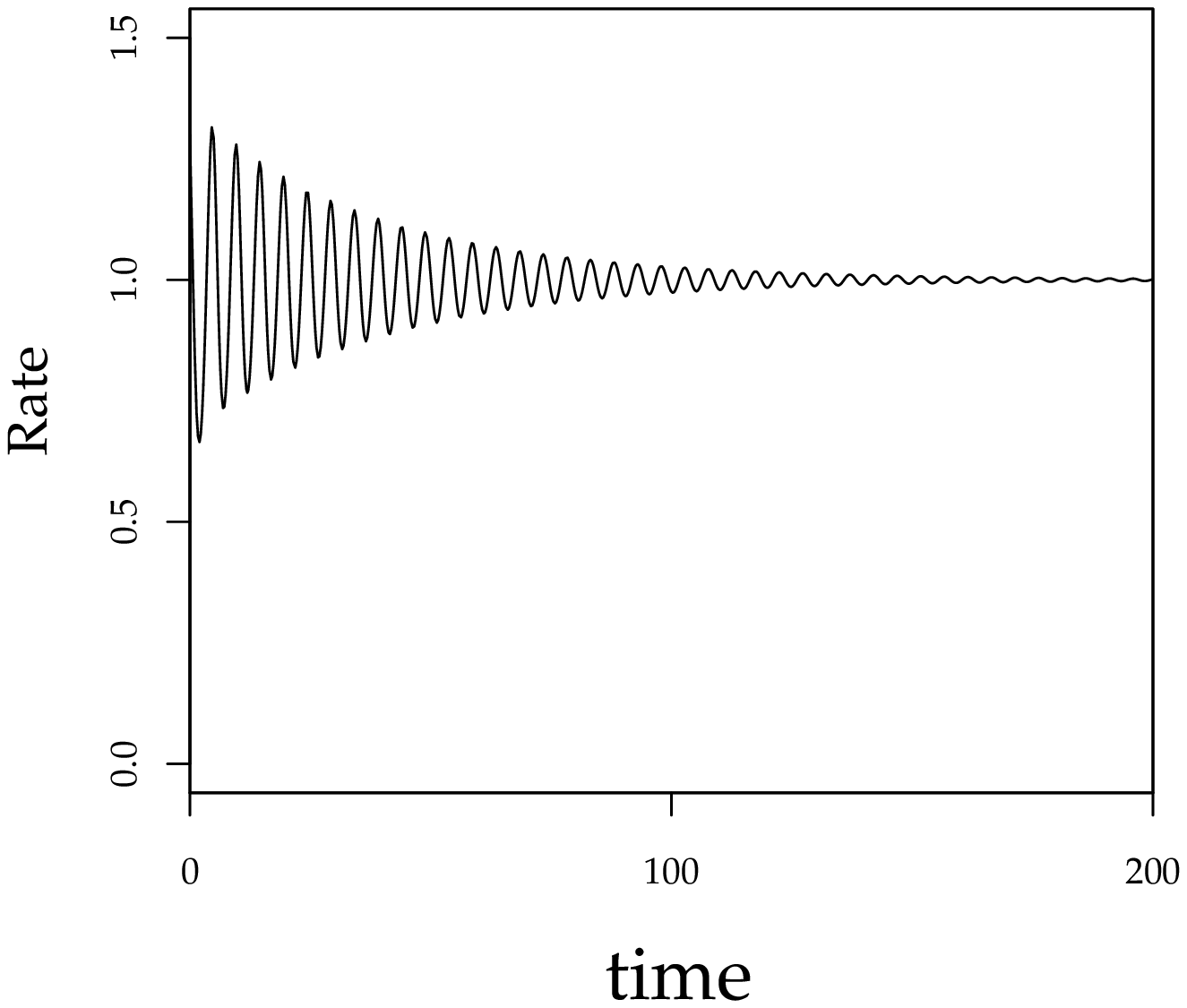}}\hspace{-5mm}
%\subfloat[ $\kappa = 0.97, R_0=2$]{\includegraphics[width=\supwdth]{tompecssuper2.eps}}\hspace{-5mm}
\subfloat[ $\kappa = 1.05$]{\includegraphics[width=\supwdth]{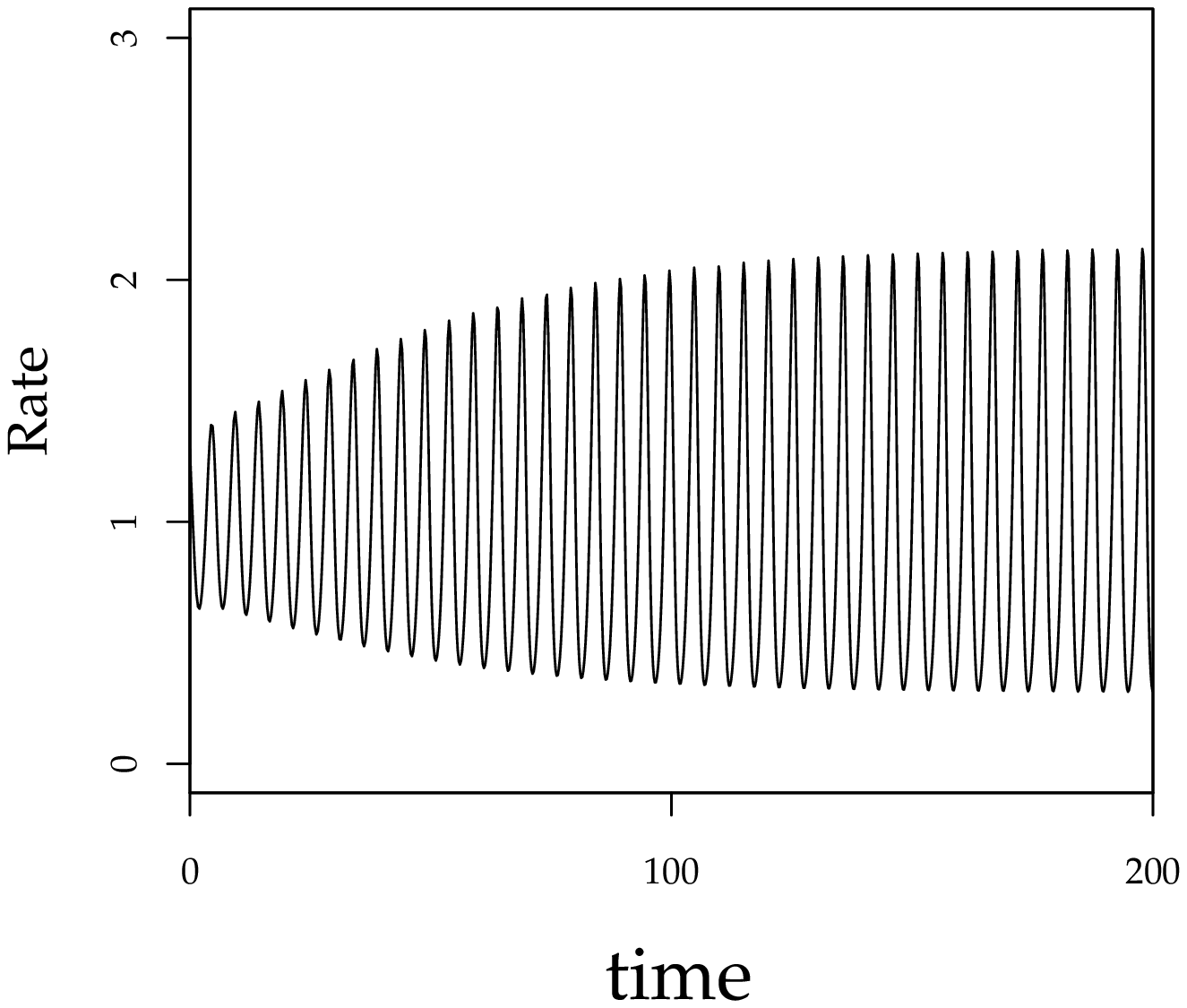}}
\end{tabular}
\caption{Numerical simulations illustrating the occurrence of a super-critical Hopf for $a=1.25$, $\beta=0.454$, $C = 1$ and $\tau = 1$.} \label{fig:sup_rcp_withq}
\end{figure}

\begin{figure}[hbtp!]
 \psfrag{60}{\hspace{-0.2cm}\small{$50000$}}
\psfrag{0}{\small{$0$}} 
 \psfrag{10}{\hspace{0.1cm}\small{$0$}}
 \psfrag{35}{\hspace{-0.2cm}\small{$25000$}}
\psfrag{20}{\hspace{-0.2cm}\small{$10000$}}
\psfrag{15}{\hspace{-0.2cm}\small{$5000$}}
\psfrag{75}{\hspace{-0.2cm}\small{$75$}}
\psfrag{150}{\hspace{-0.2cm}\small{$150$}}
 \psfrag{40}{\hspace{-0.2cm}\small{$0$}}
\psfrag{65}{\hspace{-0.2cm}\small{$25000$}}
\psfrag{90}{\hspace{-0.2cm}\small{$50000$}}
\psfrag{250}{\small{$250$}}
\psfrag{500}{\small{$500$}}
\psfrag{300}{\small{$300$}}
\psfrag{600}{\small{$600$}}
\psfrag{1000}{\small{$1000$}}
\psfrag{1500}{\small{$1500$}}
\psfrag{2000}{\small{$2000$}}
\psfrag{4000}{\small{$4000$}}
\psfrag{10000}{\small{$10000$}}
\psfrag{20000}{\small{$20000$}}
\psfrag{75}{\small{$75$}}
\psfrag{150}{\small{$150$}}
\newcommand{\hite}{3.6cm}
\centering
\begin{tabular}{cc} 
	\hspace{0.3cm}\small{Queue Size (packets)} & \hspace{0.3cm}\small{Rate (bytes/ms)} \\
	\includegraphics[height=\hite, width=0.45\textwidth, angle = 0]{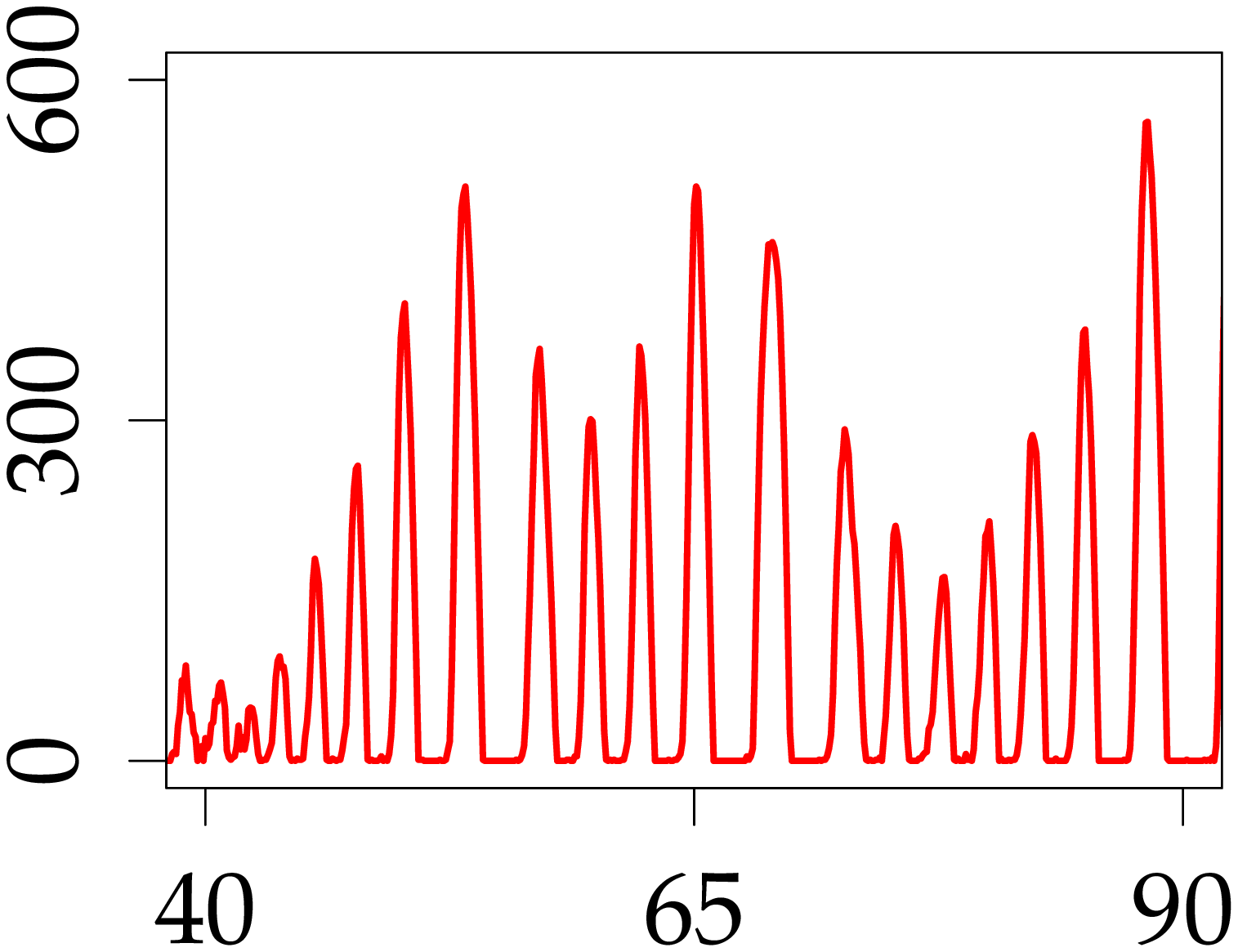} &
\includegraphics[height=\hite, width=0.45\textwidth, angle = 0]{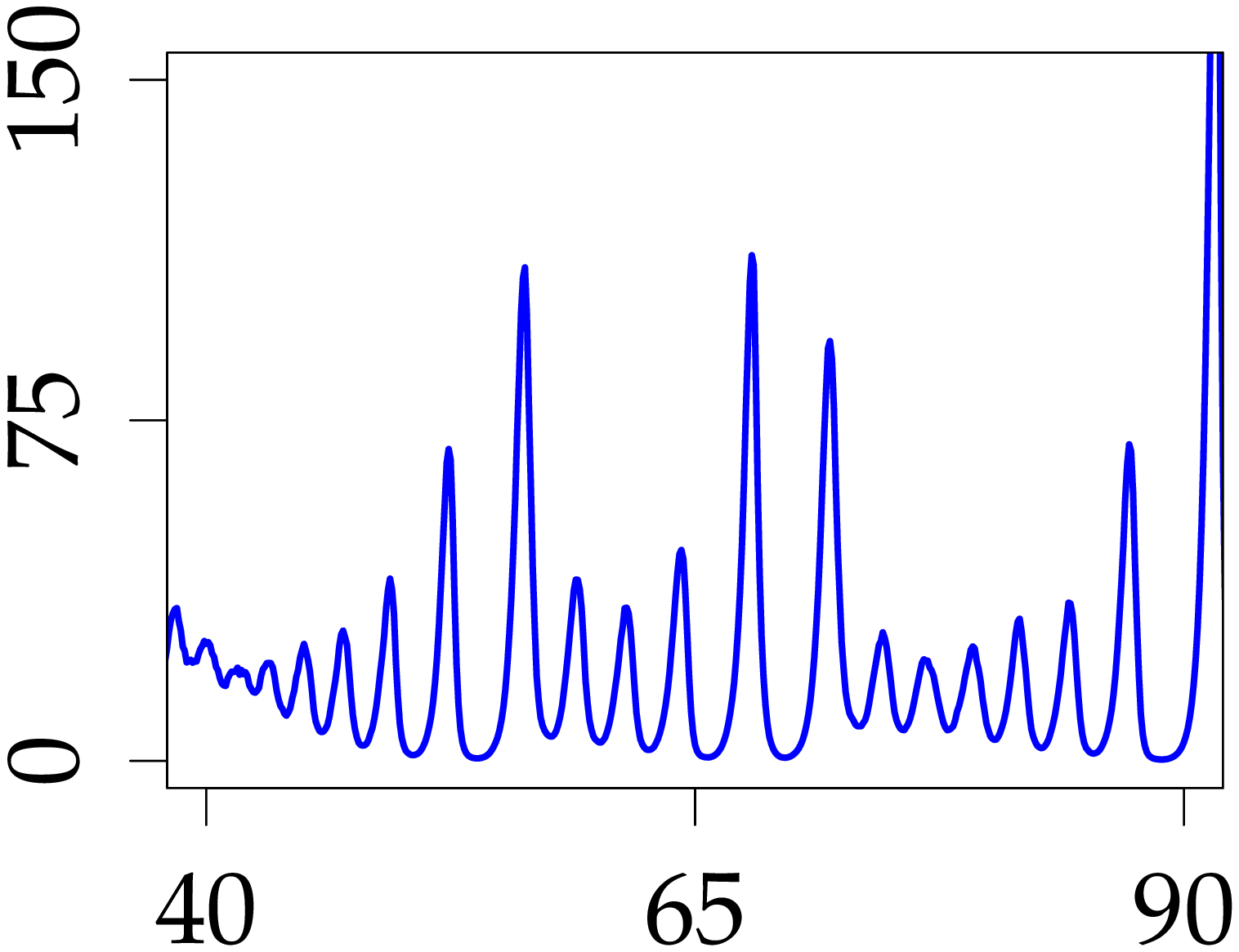} \\
 \multicolumn{2}{c}{(a)\ \small{$a = 0.8$ and $\beta = 0.55$}}\\
	\includegraphics[height=\hite, width=0.45\textwidth, angle = 0]{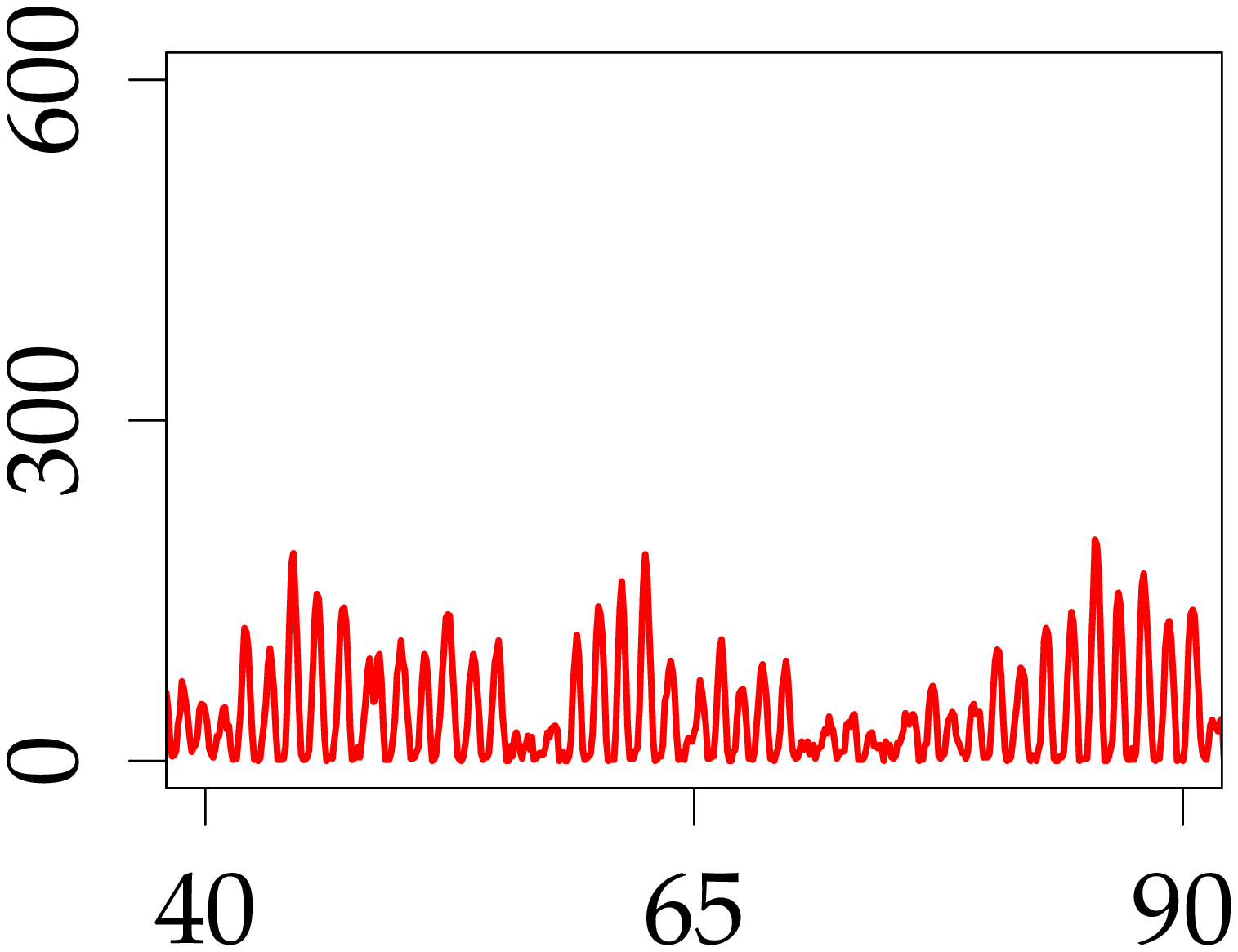} &
\includegraphics[height=\hite, width=0.45\textwidth, angle = 0]{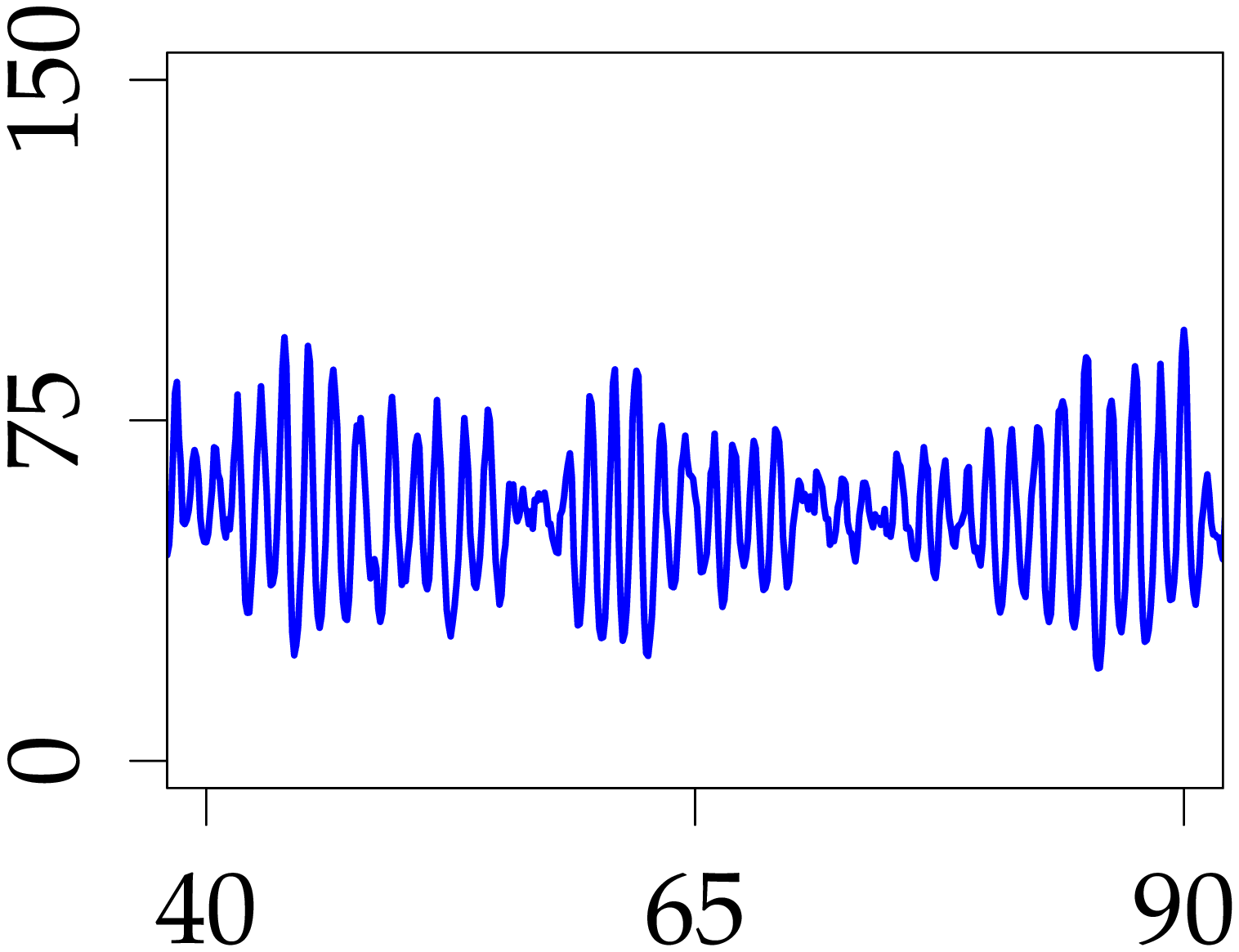} \\
 \multicolumn{2}{c}{\hspace{-2mm}(b)\ \small{$a = 1.3$ and $\beta = 0.4$}}\\
	\multicolumn{2}{c}{\small{Time, t}}\\
\end{tabular}
\caption{\small Simulation traces highlighting that the system which includes queue feedback exhibit sub-critical Hopf and super-critical Hopf for the parameter choices $a=0.8$, $\beta= 0.55$ and $a=1.3$, $\beta = 0.4$, respectively. We consider the number of flows as $100$, and each with round-trip time of $200$ time units.}
\label{fig:rcphopfbfasims}
\end{figure}

Now, we present some packet-level simulations to illustrate that the RCP with queue feedback can undergo either super-critical or sub-critical Hopf depending on the parameter values. For this simulation, we consider $C=1$ packet per unit time, number of flows = $100$ and $\tau=200$ time units. Simulation traces in  Figure \ref{fig:rcphopfbfasims} (a) and Figure \ref{fig:rcphopfbfasims} (b) show the queue size and rate for the parameter values $a = 0.8$, $\beta = 0.55$ and $a = 1.3$, $\beta = 0.4$, respectively. For $a = 0.8$ and $\beta = 0.55$, we get $\myp= 0.9 < \myp_h $, and hence the type of Hopf is expected to be sub-critical. Indeed, this is confirmed in the simulation traces in Figure \ref{fig:rcphopfbfasims} (a) which shows the emergence of large amplitude limit cycles. However, it does not lead to blow-up due to some constraints of simulators. For the parameter values $a = 1.3$ and $\beta = 0.4$, the system exhibits a super-critical Hopf which results in small amplitude stable limit cycles (see Figure \ref{fig:rcphopfbfasims} (b)). This is as expected, since the values $a = 1.3$ and $\beta = 0.4$ yield $\myp= 1.35 > \myp_h $. 
%%%%%%%%%%%%%%%%%%%%%%%%% SIMS for BIF Analysis%%%%%%%%%%%%%%%%%%%%%%%%%%%%%%%%%%%%

%So, these observations corroborate the occurrence of limit cycles due to the presence of queue size feedback.
\subsection{Without queue feedback}
% The Taylor series expansion of \eqref{eq:linear_withnoq} about its equilibrium is given by 
% \begin{equation} 
% \frac{d}{dt}r(t) = -\frac{\kappa a}{\tau} r(t-\tau) -\frac{\kappa a}{ C \tau}r(t)r(t-\tau). 
% \label{eq:nlwithoutq}
% \end{equation}
In \citep{voice2009maxminrcp}, it has been shown that the equation \eqref{eq:linear_withnoq} undergoes a local Hopf bifurcation at $\kappa=\kappa_c$, where $\kappa_c a=\pi/2.$ If the Hopf condition is just violated, the system would lose local stability via a super-critical Hopf bifurcation and the amplitude of the bifurcating limit cycles will be proportional to 
\begin{equation}
 R^{*}\sqrt{\frac{20\pi(\kappa-\kappa_c)}{3\pi-2}}.
\end{equation}
Here $R^*$ denotes the equilibrium of \eqref{eq:linear_withnoq}. It is also highlighted in \citep{voice2009maxminrcp} that equation \eqref{eq:linear_withnoq} cannot undergo a sub-critical Hopf bifurcation.\\

\textit{Discussion}: It is rather striking to find the possibility of a sub-critical Hopf bifurcation in RCP. A sub-critical Hopf would lead either to the emergence of unstable limit cycles, or to the emergence of limit cycles with a very large amplitude. Either of these outcomes is detrimental to system performance and is undesirable in engineering applications. Therefore, in the design of congestion control algorithms in which there is a possibility of violation of the conditions for local stability, an important design objective is not only to make sure that the system is stable, but also to ensure any loss of system stability produces stable limit cycles with small amplitude. Thus, the results of Hopf bifurcation analysis favors the design choice which uses only rate mismatch feedback.

\section{Condition for non-oscillatory convergence}
% We have shown that faster convergence can be achieved if we exclude the queue size feedback from the RCP. In addition to a faster rate of convergence, it is also required to have the system equilibrates without oscillations. 
Non-oscillatory convergence is a desirable characteristic in the design of dynamical systems. In this section, we derive a necessary and sufficient condition to ensure non-oscillatory convergence of RCP which uses only rate mismatch feedback.

For the system to be non-oscillatory, the eigenvalues should be negative real numbers. Recall that the characteristic equation of the RCP without queue size feedback is
\begin{equation}
\label{eq:ce_withoutq1}
\lambda + \left(\frac{a}{\tau}\right)e^{-\lambda \tau} = 0.
\end{equation}
Substituting $\lambda=-\sigma+j\omega$ in \eqref{eq:ce_withoutq1} gives 
\begin{align}
 \sigma&=\frac{a}{\tau}e^{\sigma\tau}\cos(\omega\tau), \label{eq:nocsigma}\\                         
 \omega&=\frac{a}{\tau}e^{\sigma\tau}\sin(\omega\tau). \label{eq:nocomega}
\end{align}
Solving the equations \eqref{eq:nocsigma} and \eqref{eq:nocomega} yields
\begin{equation}
 \sigma\tau=\frac{\omega\tau}{\tan(\omega\tau)}.\label{eq:nocnoq1}
\end{equation}
The right hand side of \eqref{eq:nocnoq1} is a decreasing function of $\omega$, and has a maximum value of $1$ at $\omega=0$. For the uniqueness of the solution, we require $\sigma\tau\geq 1$. Now, \eqref{eq:nocomega} can be rewritten as 
\begin{equation}
 a e^{\sigma\tau} \frac{\sin(\omega\tau)}{\omega\tau}=1.
\end{equation}
Taking the limit $\omega\rightarrow 0$ gives 
\begin{equation}
 \lim_{\omega\rightarrow 0}\ a e^{\sigma\tau} \frac{\sin(\omega\tau)}{\omega\tau}=ae^{\sigma\tau},
\end{equation}
and hence $a e^{\sigma\tau}=1$. Since $\sigma\tau\geq 1$, then $a\leq(1/e)$.
Thus, the necessary and sufficient condition for non-oscillatory convergence of RCP without queue size feedback is
\begin{equation}
a \leq \frac{1}{e}.
\end{equation}

%Put some figures of R(t) Vs t for various values of a i.e a<1/e and for a>1/e..
\begin{figure}
\psfrag{time}{\hspace{-0.1cm}  \small Time}
\psfrag{Rate}{\hspace{-0.5cm}  \small Rate, $R(t)$}
\psfrag{0}{\small{$0$}}
\psfrag{5}{\hspace{-0.05cm}\small{$5$}}
\psfrag{10}{\hspace{-0.1cm}\small{$10$}}
\psfrag{15}{\hspace{-0.1cm}\small{$15$}}
\psfrag{20}{\hspace{-0.1cm}\small{$20$}}
\psfrag{0}{\small{$0$}}
\psfrag{25}{\hspace{0cm}\small{$25$}}
\psfrag{50}{\hspace{-0.1cm}\small{$50$}}
\psfrag{100}{\hspace{-0.1cm}\small{$100$}}
\psfrag{200}{\hspace{-0.1cm}\small{$200$}}
\begin{tabular}{c}
\subfloat[ $a = 0.1, 1/e$]{\includegraphics[width=0.52\wdth]{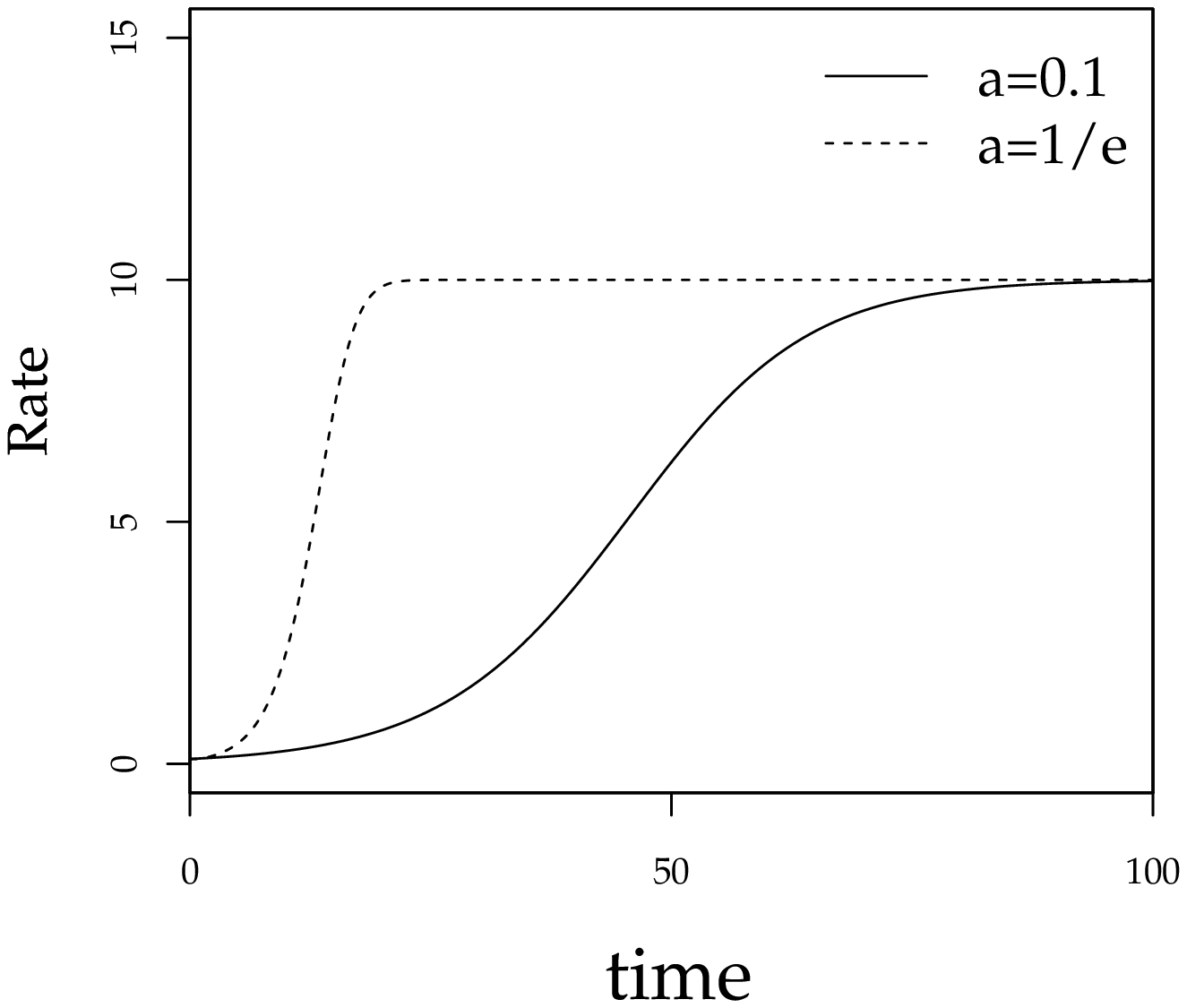}}%prev width = 2.15 in
\subfloat[ $a = 1.2$]{\includegraphics[width=0.52\wdth]{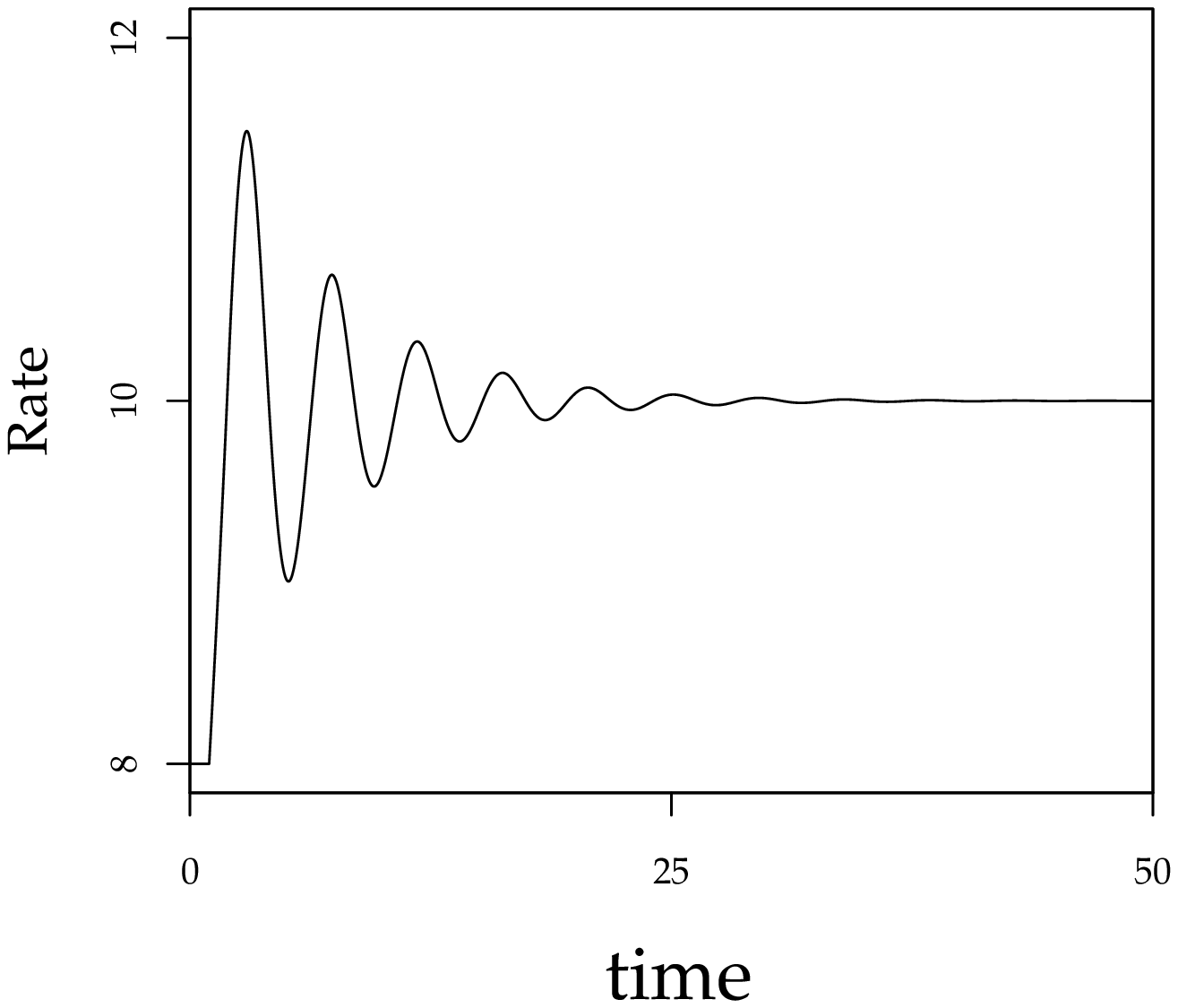}}\\
\subfloat[ $a = \pi/2$]{\includegraphics[width=0.52\wdth]{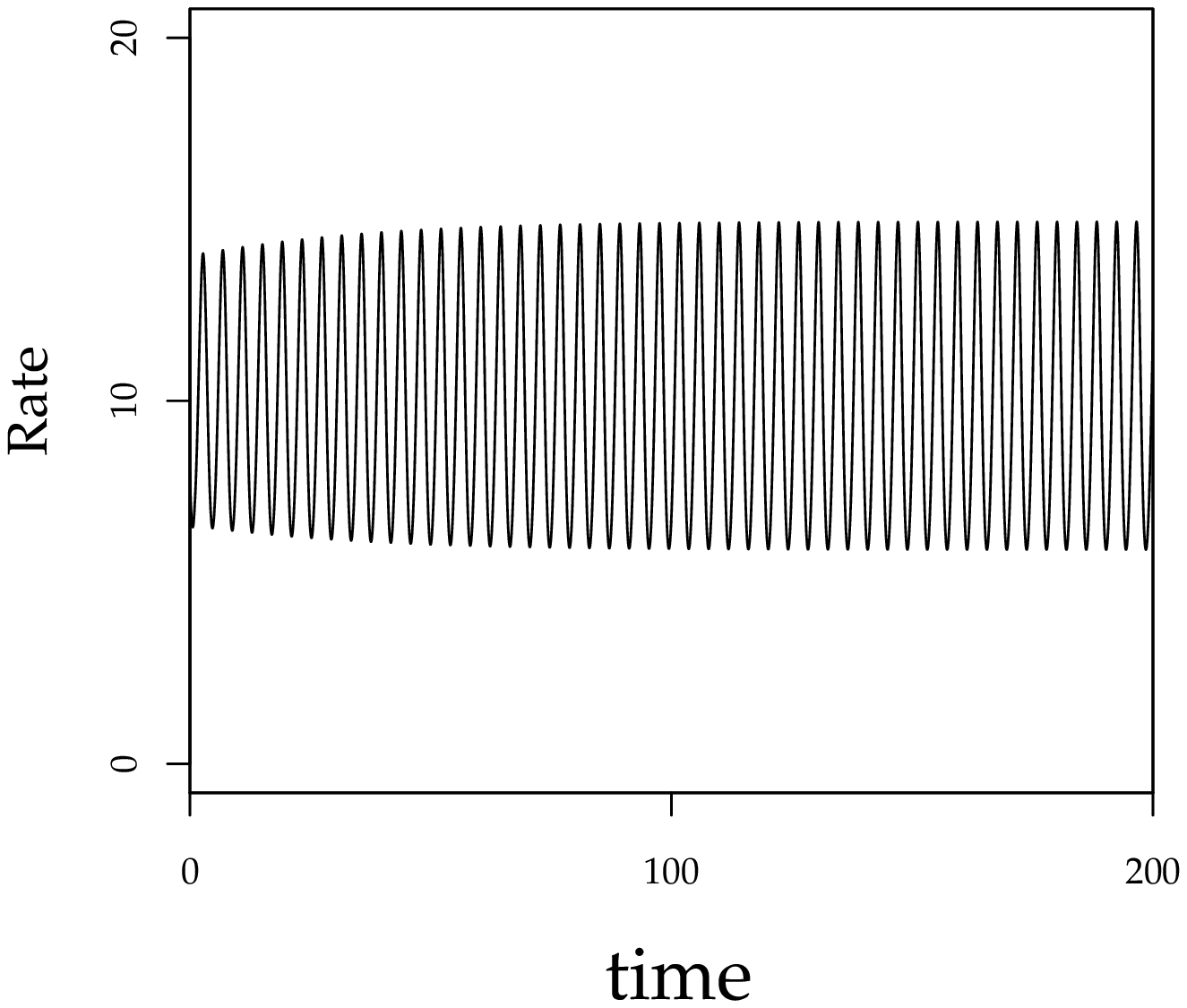}}%prev width = 2.15 in
\subfloat[ $a = 1.6$]{\includegraphics[width=0.52\wdth]{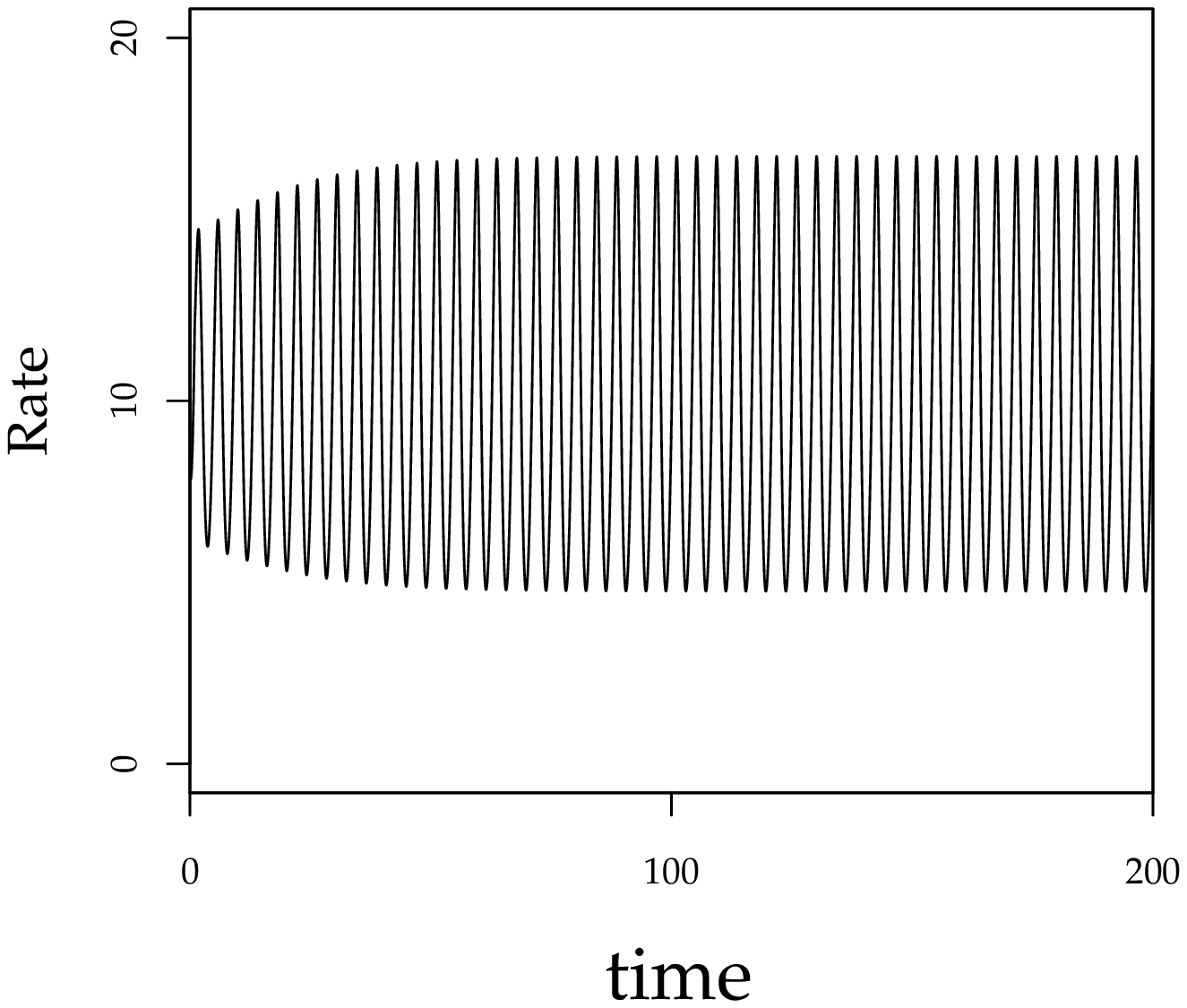}}
\end{tabular}
\caption{Numerical simulations showing the behavior of the system for various values of the protocol parameter $a$. The parameter values used are $C = 10$ and $\tau = 1$.} \label{fig:rocVsa}
\end{figure}
 
% \begin{figure}[ht!]
% \centering
% \psfrag{R}{\hspace{-0.2cm}  \small rate}
% \psfrag{a}{{\hspace{-0.8cm}parameter $a$}}
% \psfrag{5}{\hspace{0cm}\small{$5$}}
% \psfrag{10}{\hspace{0cm}\small{$10$}}
% \psfrag{15}{\hspace{0cm}\small{$15$}}
% \psfrag{20}{\small{$20$}}
% \psfrag{0}{\small{$0$}}
% \psfrag{25}{\hspace{0cm}\small{$25$}}
% \psfrag{50}{\small{$50$}}
% \psfrag{1.8000}{\small{$1.8$}}
% \psfrag{1.5000}{\small{$1.5$}}
% \psfrag{1.7000}[][]{\small{$1.7$}}
% \psfrag{1.5707}{\small{$\pi/2$}}
% \includegraphics[width=3.6in]{rcp_bfd_cep1.eps}
% \caption{Bifurcation diagram showing the existence of super-critical Hopf bifurcation as the protocol parameter increases beyond $\pi/2$. The parameter values used are $C=10$ and $\tau=1$.}
% \label{fig:bfd1}
% \end{figure}
After analyzing the condition for non-oscillatory convergence, and the rate of convergence, it can be deduced that the rate of convergence is maximum at the boundary of non-oscillatory regime, i.e., $a=1/e$. Thus, the analytical results  reveals that the optimal value of the protocol parameter $a$ is $1/e$. The system becomes oscillatory when $a>1/e$, and hence the rate of convergence decreases. At $a=\pi/2$, the convergence rate is zero and the reason behind is that the system transits into unstable regime via a Hopf bifurcation at $a=\pi/2$.

To validate the theoretical results of non-oscillatory convergence, numerical simulations obtained using XPPAUT are shown in Figure \ref{fig:rocVsa}. As can be observed in Figure \ref{fig:rocVsa} (a), for $a=0.1$ $(<1/e)$, the system shows over-damped behavior, i.e., reaches equilibrium without oscillating. At $a=1/e$, the system reaches equilibrium as quickly as possible without any oscillations. For $a \in(1/e, \pi/2)$, the system behaves in an under-damped manner, i.e., existence of convergent oscillations (see Figure \ref{fig:rocVsa} (b)). As shown in Figure \ref{fig:rocVsa} (c) and Figure \ref{fig:rocVsa} (d), the system loses stability at $a=\pi/2$, and the amplitude of the undamped oscillations increases as $a$ increases beyond $\pi/2$.

% Also, the rate of convergence decreases with the increase in RTT (see Figure Figure \ref{fig:rocVstau}). The numerical simulations shown in Figure ~Figure \ref{fig:rocVsa} and Figure \ref{fig:rocVstau} are done using computing software XPPAUT.
% The bifurcation diagram drawn using DDE-Biftool is shown in the Figure Figure \ref{fig:bfd1}. It clearly shows the existence of local super-critical Hopf as the protocol parameter $a$ increases beyond $\pi/2$. The phase portraits (done with MATLAB) shown in Figure Figure \ref{fig:phaseportrait}, clearly exhibits the emergence of limit cycles from a stable fixed point, as $a$ increases.
The dependence of system behavior on the protocol parameter value is summarized in \eqref{tab:roc}.
% \begin{table}[!htbp]
% \caption{Effect of the value of protocol parameter on the system behavior.}\label{tab:roc}
% \renewcommand{\arraystretch}{2.25}
% \centering
% \begin{tabular}{ll}
% \hline
% $\text{Parameter range} \qquad \qquad \qquad \qquad \qquad$  & $\text{System behavior} $\\
% \hline
% %\vspace{-5mm}\\
% $a \in (0,1/e]$\ \   & stable and non-oscillatory\\  
% $a \in (1/e , \pi/2)$\ \   & stable and oscillatory\\
% $a\geq\pi/2$\ \   &  unstable\\
% \hline
% \end{tabular}
% \\
% \end{table}

\begin{table}
\caption{Effect of the value of protocol parameter on the system behavior. At $a=1/e$, the system reaches equilibrium quickly without any oscillations.}\label{tab:roc}
  \begin{minipage}{\columnwidth}
\centering
\resizebox{0.55\columnwidth}{!}{%
\begin{tabular}{ll}
  \toprule
    Parameter range&System behavior\\
    \midrule
    $a \in (0,1/e]$  & stable and non-oscillatory\\  
$a \in (1/e , \pi/2)$  & stable and oscillatory\\
$a\geq\pi/2$  &  unstable\\
  \bottomrule
  \end{tabular}
  }
%\end{center}
\end{minipage}
\end{table}%

\section{Conclusions}
The design of explicit congestion control protocols is an important research topic among the networking research community. The Rate Control Protocol (RCP) is a well-known explicit feedback algorithm that utilizes router feedback based on rate mismatch and queue size. However, it is currently an open question if the protocol definition should include two forms of feedback; i.e., both rate mismatch and queue feedback. In this paper, we developed a better understanding of this design choice, using tools from control and bifurcation theory. In particular, we considered stability, convergence and the impact of the loss of local stability on system dynamics.      

Local stability analysis was conducted both in the presence and absence of queue size feedback and established necessary and sufficient conditions to ensure local stability. This enabled us to determine bounds on the protocol parameters to guarantee the local stability of the system. The rate of convergence analysis highlighted that the queue term should be excluded from the protocol to improve the convergence rate. Also, a necessary and sufficient condition that guarantees non-oscillatory convergence was derived for RCP where the queue size term is absent. This condition enabled us to suggest an optimal value of the protocol parameter.    

Further, we also showed that in the presence of queue feedback, the system readily loses its stability via a Hopf bifurcation, as the bifurcation parameter varies. For our analyses, a dimensionless exogenous parameter was used as the bifurcation parameter. The phenomenon of Hopf bifurcation would result in the onset of limit cycles which make it harder to control the queue size. We analyzed the Hopf bifurcation properties by applying Poincar\'{e} normal forms and the center manifold theorem. We showed that, for some parameter values, the presence of queue feedback in the RCP model leads to a sub-critical Hopf bifurcation and the emergence of unstable limit cycles. But, in the absence of queue feedback, the Hopf bifurcation is super-critical, and the bifurcating periodic solutions are asymptotically orbitally stable. In essence, all our insights revealed that the performance of the RCP could be improved by removing the queue size term from the protocol definition. Thus, based on our current analysis, we suggest it is preferable to go with the design choice that uses only rate mismatch feedback. However, this is a critical design consideration, and additional analysis and experimental results would be needed to arrive at a comprehensive understanding. 

A natural extension of this study would be to validate the analytical insights through hardware experiments. In the real network environment, flows arrive and depart dynamically, and also can have heterogeneous round-trip times. Thus, future research should also include an in-depth analysis of the impact of queue feedback on the performance of the system with multiple time delays, and for a variety of flow arrival and departure patterns.

% 
% \section*{Acknowledgments}
% We would like to acknowledge the assistance of volunteers in putting
% together this example manuscript and supplement.

% \bibliographystyle{siamplain}
% \bibliography{references}

\end{document}